\documentclass[dvips]{acta1}
\usepackage{supertabular,lscape,epsfig}
\usepackage{amssymb}
\usepackage{amsmath}
\DeclareSymbolFont{ppa}{OT1}{ppl}{m}{it}
\DeclareMathSymbol{\vv}{\mathalpha}{ppa}{'166}

\newfont{\hb}{rphvb at 10pt}%bezszeryfowe pó³grube
\newfont{\hbo}{rphvbo at 10pt}%bezszeryfowe pó³grube kursywa
\newfont{\bitt}{rptmbi at 12pt}%pó³gruba kursywa (tytu³ artyku³u)
\newfont{\bits}{rptmbi at 11pt}%pó³gruba kursywa (tytu³y rozdzia³ów)

\SetPages{133}{151}

\SetVol{62}{2012}

\begin{document}

%Zwarte naglowki, jeden wiersz, appendix
\newcommand{\TabApp}[2]{\begin{center}\parbox[t]{#1}{\centerline{
  {\bf Appendix}}
  \vskip2mm
  \centerline{\small {\spaceskip 2pt plus 1pt minus 1pt T a b l e}
  \refstepcounter{table}\thetable}
  \vskip2mm
  \centerline{\footnotesize #2}}
  \vskip3mm
\end{center}}

%Zwarte naglowki, jeden wiersz
\newcommand{\TabCapp}[2]{\begin{center}\parbox[t]{#1}{\centerline{
  \small {\spaceskip 2pt plus 1pt minus 1pt T a b l e}
  \refstepcounter{table}\thetable}
  \vskip2mm
  \centerline{\footnotesize #2}}
  \vskip3mm
\end{center}}

%Zwarte naglowki, dwa wiersze
\newcommand{\TTabCap}[3]{\begin{center}\parbox[t]{#1}{\centerline{
  \small {\spaceskip 2pt plus 1pt minus 1pt T a b l e}
  \refstepcounter{table}\thetable}
  \vskip2mm
  \centerline{\footnotesize #2}
  \centerline{\footnotesize #3}}
  \vskip1mm
\end{center}}

%Zwarte naglowki, jeden wiersz, appendix
\newcommand{\MakeTableApp}[4]{\begin{table}[p]\TabApp{#2}{#3}
  \begin{center} \TableFont \begin{tabular}{#1} #4 
  \end{tabular}\end{center}\end{table}}

%Zwarte naglowki, jeden wiersz
\newcommand{\MakeTableSepp}[4]{\begin{table}[p]\TabCapp{#2}{#3}
  \begin{center} \TableFont \begin{tabular}{#1} #4 
  \end{tabular}\end{center}\end{table}}

%Zwarte naglowki, jeden wiersz
\newcommand{\MakeTableee}[4]{\begin{table}[htb]\TabCapp{#2}{#3}
  \begin{center} \TableFont \begin{tabular}{#1} #4
  \end{tabular}\end{center}\end{table}}

%Zwarte naglowki, dwa wiersze
\newcommand{\MakeTablee}[5]{\begin{table}[htb]\TTabCap{#2}{#3}{#4}
  \begin{center} \TableFont \begin{tabular}{#1} #5 
  \end{tabular}\end{center}\end{table}}

%{\it Acta Astronomica Archive}
%\parskip=0pt \itemsep=1mm \setlength{\itemsep}{0.4mm}\setlength{\parindent}{-1em} \setlength{\itemindent}{-1em} - po \begin{itemize} - wszystko
%FWHM, PSF, S/N - proste, 
%MgII, H$\alpha$
%rms, rhs, sd - kursywa
%{\sc DAOPhot}
%{\sf files}
%Galactic wszystko (bulge, center, plane, disk, coordinates, latitudes...)
%Cepheids
%type~ Cepheids, Population~II Cepheids
%a.u.
\newfont{\bb}{ptmbi8t at 12pt}
\newfont{\bbb}{cmbxti10}
\newfont{\bbbb}{cmbxti10 at 9pt}
\newcommand{\uprule}{\rule{0pt}{2.5ex}}
\newcommand{\douprule}{\rule[-2ex]{0pt}{4.5ex}}
\newcommand{\dorule}{\rule[-2ex]{0pt}{2ex}}
\def\thefootnote{\fnsymbol{footnote}}
\begin{Titlepage}
\Title{The Optical Gravitational Lensing Experiment.
Optical Counterparts to the X-ray Sources in the Galactic Bulge
\footnote{Based on observations obtained with the 1.3~m Warsaw telescope
at the Las Campanas Observatory of the Carnegie Institution for
Science.}}
\Author{A.~~U~d~a~l~s~k~i$^1$, ~~K.~~K~o~w~a~l~c~z~y~k$^1$, ~~I.~~S~o~s~z~y~ñ~s~k~i$^1$,~~R.~~P~o~l~e~s~k~i$^1$,
M.\,K.~~S~z~y~m~a~{ñ}~s~k~i$^1$, ~~M.~~K~u~b~i~a~k$^1$, 
~~G.~~P~i~e~t~r~z~y~ñ~s~k~i$^{1,2}$, ~~S.~~K~o~z~³~o~w~s~k~i$^1$, 
~~P.~~P~i~e~t~r~u~k~o~w~i~c~z$^1$,~~K.~~U~l~a~c~z~y~k$^1$,
~~J.~~S~k~o~w~r~o~n$^{3,1}$,~~and~~£.~~W~y~r~z~y~k~o~w~s~k~i$^1$}
{$^1$Warsaw University Observatory, Al.~Ujazdowskie~4, 00-478~Warszawa, Poland\\
e-mail:
(udalski,kkowalczyk,soszynsk,rpoleski,msz,mk,pietrzyn,simkoz,pietruk, 
kulaczyk,jskowron,wyrzykow)@astrouw.edu.pl\\
$^2$ Universidad de Concepción, Departamento de Astronomia, Casilla 160--C, Concepción, Chile\\
$^3$ Department of Astronomy, Ohio State University, 140 W.~18th Ave., Columbus, OH~43210, USA}
\Received{July 20, 2012}
\end{Titlepage}

\Abstract{We present a sample of 209 variable objects -- very likely
optical counterparts to the X-ray sources detected in the direction of the
Galactic center by the Galactic Bulge Survey (GBS) carried out with the
Chandra satellite. The variable sources were found in the databases of the
OGLE long term survey monitoring regularly the Galactic bulge since
1992. The counterpart candidates were searched based on the X-ray source
position in the radius of 3\zdot\arcs9.

Optical light curves of the candidates comprise a full variety of
variability types: spotted stars, pulsating red giants (potentially
secondary stars of symbiotic variables), cataclysmic variables,
eclipsing binary systems, irregular non-periodic objects including an
AGN (GRS 1734--292). 

Additionally, we find that positions of 19 non-variable stars brighter
than 16.5~mag in the OGLE databases are so well aligned with the X-ray
positions ($<0\zdot\arcs75$) that these objects are also likely optical
counterparts to the GBS X-ray sources.

We provide the OGLE astrometric and photometric information for all
selected objects and their preliminary classifications. Photometry
of the candidates is available from the OGLE Internet archive.}{X-ray:
stars -- X-ray: binaries -- Stars: activity -- binaries: symbiotic --
novae, cataclysmic variables -- Galaxies: Seyfert}

\Section{Introduction}
Recent years have been witnessing an enormous progress in the X-ray
astronomy thanks to many satellite missions with continuously more
sensitive state-of-the-art detectors. The number of detected X-ray
sources increased rapidly. For example, the surveys of the Magellanic
Clouds led to the discovery of tens of new X-ray pulsars (Coe \etal 2005)
and AGNs (Koz³owski \etal 2012), Galactic center survey discovered more
than thousand new X-ray sources toward the Galactic center (Jonker \etal
2011), the XMM-Newton catalog of serendipitous X-ray sources includes
almost 200\,000 objects (Watson \etal 2009),

Proper classification and then interpretation of an X-ray source usually
requires additional observations in other wavelength bands, usually in the
optical or IR range. Therefore, it is crucial to find the X-ray source
counterparts in these bands, if the data exist.
  
The accuracy of positions of X-ray sources detected by the modern
instruments is comparable to that obtained in the optical range. This in
principle can make the cross-identification of the optical counterparts
relatively straightforward. Indeed, it is often true in empty stellar
fields, where the X-ray position unambiguously points to an object. The
situation is, however, much more difficult in dense stellar fields
such as the Magellanic Clouds or Galactic center, where one can detect
even several dozen objects within the X-ray position error box. In such
cases positive cross-identification can often be non-trivial.

On the other hand, the photometric variability of an object located close to
the X-ray source position can be a very good indicator of its close
relation with the detected X-ray source. It is well known that many
different types of variable objects emit X-rays, for example, all types of
cataclysmic variables, low and high mass X-ray binaries, chromospherically
active and spotted stars, long period variables etc. Thus, the detection of
optical variability in an object located in the X-ray position error box
practically proves correctness of the cross-identification.

What is more important, variability provides many additional information on
the X-ray system for its proper classification and detailed studies of its
structure. For example, the detected periodicities can be interpreted as
related to orbital or rotational periods. Observed flares or flickering can
provide information on the rapid phenomena occurring in accreting systems
or on stellar chromospherical activity.

Here, we present the results of our search for variable objects located at
the positions of the X-ray sources detected by the Galactic Bulge Survey
(GBS, Jonker \etal 2011). Altogether, we have found 209 variable optical
objects that can be considered as the optical counterparts to X-ray
sources detected by the GBS survey. For all of them we provide their basic
classification and light curve parameters. The photometric data are
available from the OGLE Internet archive (see Section~5).

\Section{Observational Data}
The X-ray sources analyzed in this paper were detected by the GBS project
(Jonker \etal 2011). They list 1234 sources detected by the Chandra
satellite in two regions located symmetrically along the Galactic plane:
$-3\arcd<l<3\arcd$, $1\arcd<|b|<2\arcd$, \ie covering 12 square degrees
near the Galactic center and avoiding the highest extinction regions close
to the Galactic plane. The GBS catalog provides the positions of the
detected sources with typical accuracy of about 0\zdot\arcs25 in right
ascension and declination, although in rare single cases the accuracy of
position can be worse than 1\zdot\arcs5.
  
The main optical dataset comes from the fourth phase of the Optical
Gravitational Lensing Experiment (OGLE-IV). Observations were collected
between March 2010 and June 2012 (and the data are still being collected
as the OGLE-IV project continues). The 1.3-m Warsaw Telescope located at
Las Campanas Observatory, Chile, operated by the Carnegie Institution
for Science was used with the OGLE-IV 32-chip mosaic camera covering
approximately 1.4 square degrees on the sky with the scale of 0.26
arcsec/pixel.

As the OGLE-IV survey focuses on variability, observations have been
carried out mostly in the {\it I}-band filter. About 10\% images are
taken in the {\it V}-band to secure color information. The cadence of
observations depends on the field. In the regions where the
gravitational microlensing phenomena (main target of the OGLE-IV survey
in the Galactic bulge) occur most frequently the cadence is very high
reaching one observation per 18 minutes. Other fields are observed less
frequently but typically not less than one observation per one/two
nights during the mid observing season is obtained. The OGLE-IV light
curves presented here have from more than 5000 to 100 epochs in the
highest and lowest cadence fields, respectively.

\begin{figure}[htb]
\centerline{\includegraphics[width=12.7cm]{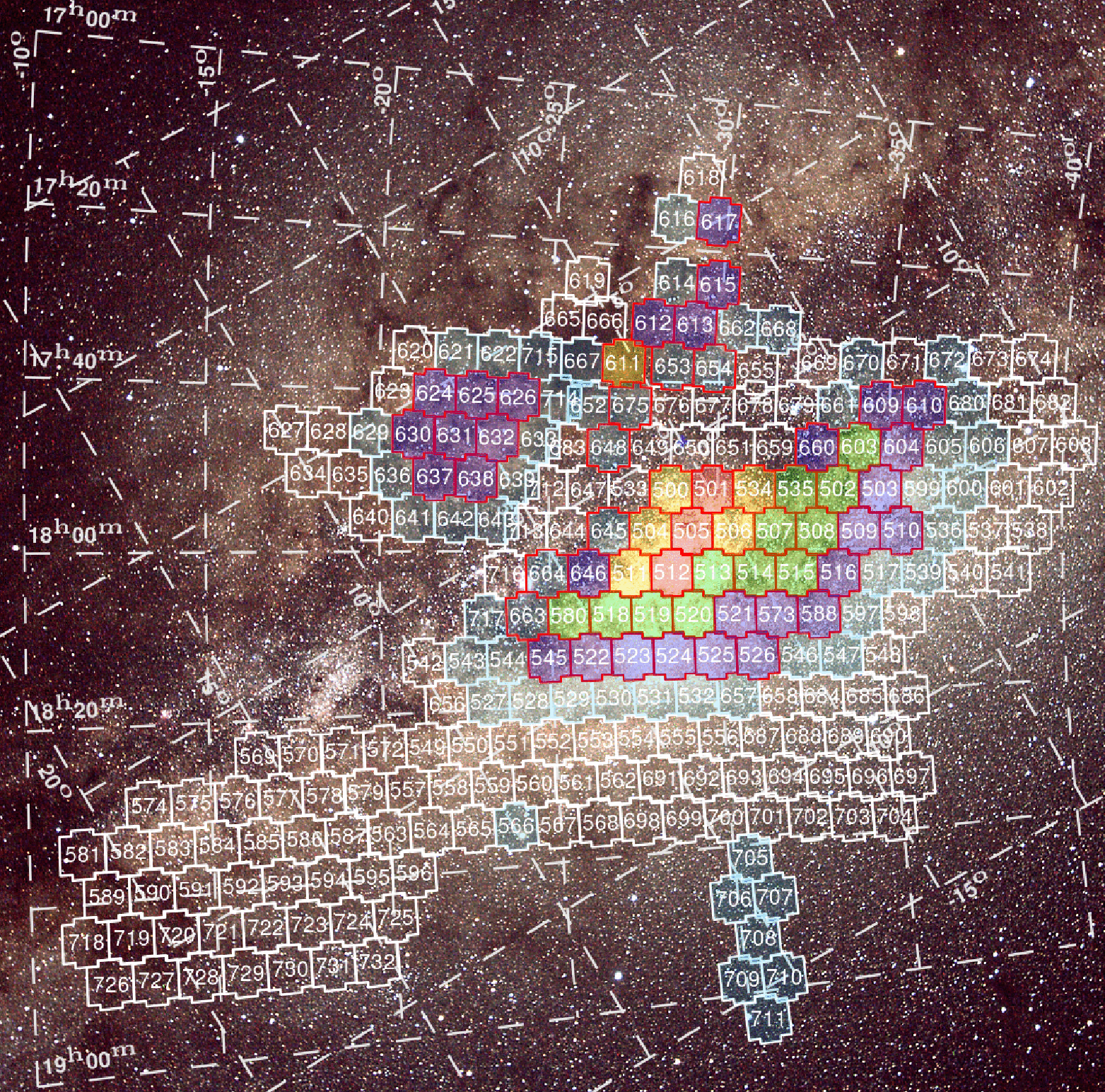}}
\vskip6pt
\FigCap{Map of the fields observed by the OGLE-IV survey in the Galactic
bulge colored according to the cadence of observations. Dashed white lines
mark the equatorial and Galactic coordinate grids (every 5\arcd in $l$ and
$b$). Red field borders indicate fields with photometry available for
analysis presented in this paper.}
\end{figure}
Fig.~1 shows the map of the Galactic center with the coverage by the
OGLE-IV survey. It can be compared with the map of GBS survey coverage
(Fig.~1, Jonker \etal 2011).

The exposure time of {\it I}-band images was set to 100~s and to 150~s for
the {\it V}-band frames. Because of very high stellar density, observations
of the Galactic bulge are conducted only during good seeing (less than
1\zdot\arcs8) and transparency conditions.

Photometry of all objects located in the OGLE-IV Galactic center fields is
derived with the image subtraction method (Alard and Lupton 1998, Alard
2000, Wozniak 2000). The data pipeline is based on slightly modified
OGLE-III data pipeline (Udalski 2003). At the time of this analysis
photometry of the fields marked with the red borders in Fig.~1 was
accessible. The databases of these 58 OGLE-IV fields contain photometry of
about 250 million stars. OGLE-IV photometry has so far been only roughly
calibrated and the uncertainty of the zero point may reach 0.1--0.15~mag.

Astrometry of the OGLE-IV fields was derived in the identical way as for
the OGLE-III photometric maps (Szymañski \etal 2011) and tied to the 2MASS
survey coordinate system. Comparison of the positions obtained for
thousands of cross-identified stars in OGLE-III and OGLE-IV images
indicates the internal accuracy of the astrometry at the 0\zdot\arcs1
level.

OGLE-IV observations cover the largest area overlapping with the sky
surveyed by the GBS compared to the previous OGLE phases. However,
because of some small ``dead'' regions on the sky caused by the gaps
between detectors of the OGLE-IV mosaic camera and the lack of OGLE-IV
photometric databases from a few fields closest to the Galactic plane
which still await calibration, supplementary optical data from the
OGLE-II (Udalski, Kubiak and Szymañski 1997) and OGLE-III (Udalski 2003)
phases were also analyzed. These two datasets were also collected with
the 1.3-m Warsaw telescope at Las Campanas Observatory but with previous
generation CCD cameras. OGLE-II observations were conducted in the years
1997-2000 and OGLE-III 2001-2009 with typical cadence of one/two
observations per night. It is worth noting that while the bright
unsaturated limit of the OGLE-III and OGLE-IV photometry is
approximately similar, OGLE-II data cover up to 1--1.5~mag brighter
objects.
\vspace*{-9pt}
\Section{Search for the Optical Counterparts to the X-ray Sources}
\vspace*{-5pt}
Search for optical counterparts to the GBS survey X-ray sources was
performed in a few steps. First, based on the X-ray positions we
selected a subsample of X-ray sources that fall into the regions of the
sky for which OGLE-IV databases are currently available. From 1234
sources listed in the GBS catalog 836 passed this cut. However, a subset
of this sample cannot be searched for variability. Some of the X-ray
sources are located in the gaps between the CCD detectors of the OGLE-IV
mosaic camera on the reference images, some are located so close to the
edges of the observed area that due to imperfections of the telescope
pointing the number of collected observations is too small for
variability analysis. After removing such X-ray sources, we were left
with a total of 782 objects for further analysis.

In the second step, we removed from the sample all cases where the X-ray
position closely points to a bright object which is overexposed on the
OGLE-IV reference image and thus its photometry is unavailable. In most
cases the coincidence of the X-ray and optical positions is so good that
these stars are very likely the optical counterparts to the X-ray
sources. Photometry from shallow surveys like the ASAS survey (Pojmañski
2002) should be checked to search for variability in this subsample.
There were 87 such cases in our OGLE-IV sample and they were removed in
this step.

Finally, we retrieved the OGLE-IV photometry of all objects located
within the conservative radius of 15 pixels (3\zdot\arcs9) around the
X-ray position for all the remaining sources. The size of the
circle is well over $3\sigma$ X-ray position error for the vast majority
of the analyzed X-ray sources. Depending on the stellar density of the
field, from a few to several dozen objects were extracted inside each
circle.

All extracted stars were searched for optical variability. First we
performed a period search to extract periodic variables. We used the {\sc
Fnpeaks} code kindly provided by Z.~Ko³aczkowski and, independently, the
code based on {\sc AoV} algorithm (Schwarzenberg-Czerny 1989). After the
determination of tentative periodicities all these objects were visually
inspected and the periods were refined, if necessary. Next, all the
non-periodic stars were also inspected visually to look for irregular
variable stars. Finally, the positions of all selected variable stars were
double checked on the reference images and subtracted images to verify
correct identification of each variable object. It is well known that close
neighbors blended with a variable star can mimic its variability.

205 variable sources were found within the 3\zdot\arcs9 radius of the
positions of X-ray sources detected by the GBS project in the OGLE-IV
dataset. These objects are very likely their optical counterparts.

Additionally, we repeated the same procedure for the OGLE-III and
OGLE-II datasets. Positions of 114 and 58 GBS X-ray sources point to
OGLE-III and OGLE-II fields, respectively. Eleven  additional X-ray
sources could be checked in the OGLE-III dataset including nine objects
located outside currently available OGLE-IV fields and two objects in
the OGLE-IV ``dead'' zones. No variable objects were found, however, in
the searched areas. For 30 variable objects selected in the OGLE-IV data
search OGLE-III photometry was retrieved extending the photometric
coverage of these candidates to over a decade.

Six X-ray sources not covered by the OGLE-IV data as well as two sources
from OGLE-IV ``dead'' zones were checked in the OGLE-II photometric
databases. One new variable was cross-identified with a GBS X-ray
source. For 18 variable objects found in the OGLE-IV data search OGLE-II
light curves could also be retrieved.

Additionally, because the OGLE-II photometry has brighter saturation
level all saturated in OGLE-IV potential counterparts were rechecked in
the OGLE-II databases (nine objects). Three new variables were found
among this bright stars sample increasing the total number of variable
objects -- OGLE optical counterpart candidates to GBS X-ray sources --
to 209.
\vspace*{9pt}
\Section{Discussion}
\vspace*{5pt}
We performed tentative classification of each discovered variable star
based on its photometric behavior. Initially, the variables were divided
into two broad categories: those with periodic and irregular
variability. From the former group a few sub-categories, sometimes
overlapping, were then assigned.

The most numerous sub-group -- 81 objects -- consists of spotted stars
with characteristic pattern of photometric behavior resembling RS~CVn
type stars. In most cases, we register only the variability caused by
spots (rotation, almost synchronized with orbital period in the binary
systems) and their slow migration on the stellar surface. However, the
classical eclipsing RS~CVn type stars are also present in our sample. 
Typical light curves of spotted stars are shown in Fig.~2.
\begin{figure}[p] 
\hglue-7mm{\includegraphics[width=13cm, bb=100 125 505 710]{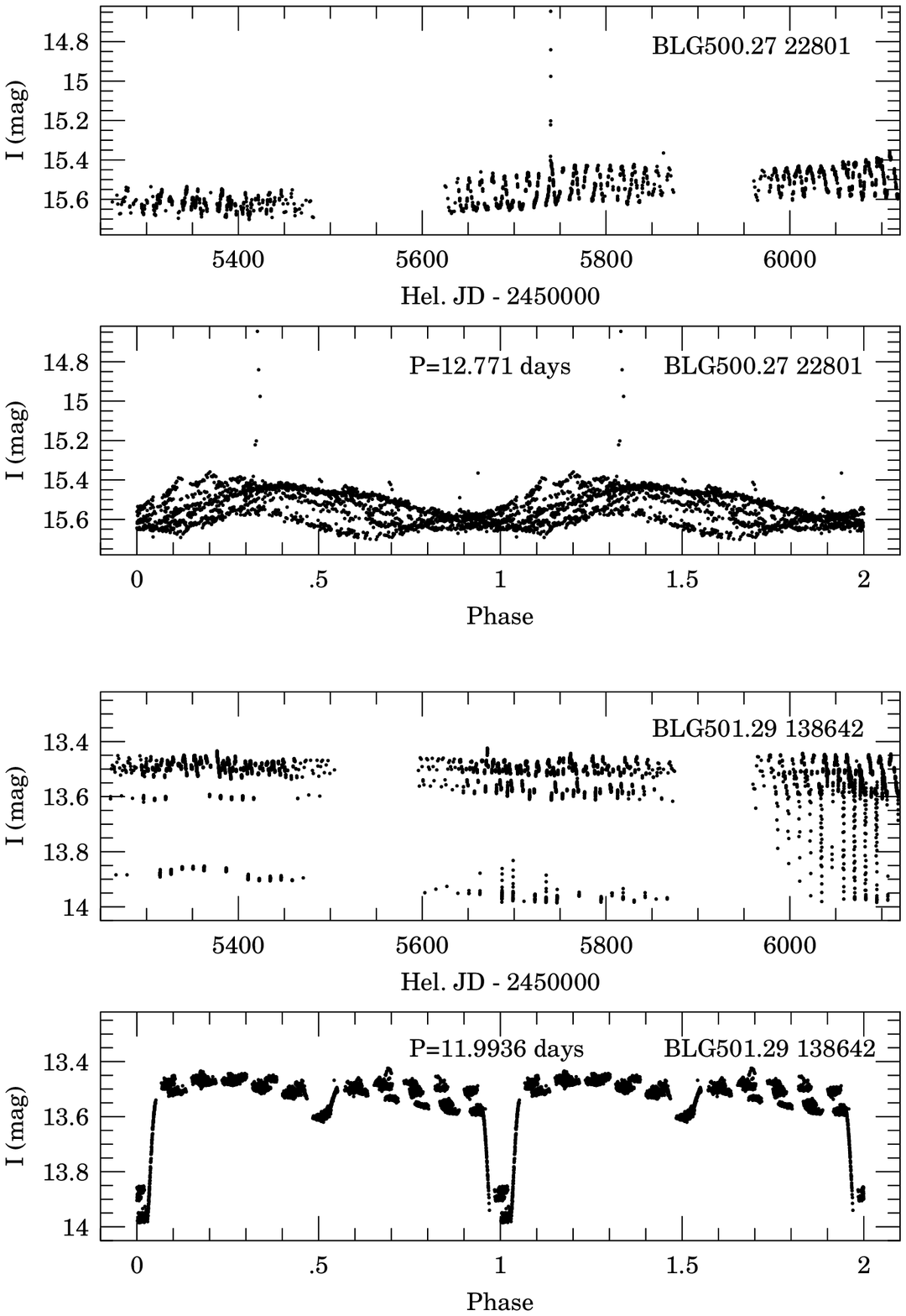}} 
\FigCap{Light curves of spotted stars. Two {\it upper panels} show time
series and phased with the period 12.771~days light curve of the optical
counterpart candidate (BLG500.27 22801) to the GBS \#29 X-ray source.
Two {\it bottom panels} show the light curves of the eclipsing system
BLG501.29 138642 ($P=11.9936$~days) -- a counterpart to the GBS \#701
X-ray source.} 
\end{figure}

The second group consists of pulsating red giants (OGLE Small Amplitude
Red Giants, OSARGs, SemiRegular Variables, SRVs, Miras - \cf Soszyñski
\etal 2011b) -- 25 objects. Such stars can be members of X-ray binary
systems with a compact companion and accreting processes. Fig.~3
presents typical light curves of objects from this sub-category.

\begin{figure}[p]
\hglue-7mm{\includegraphics[width=13cm, bb=100 125 505 710]{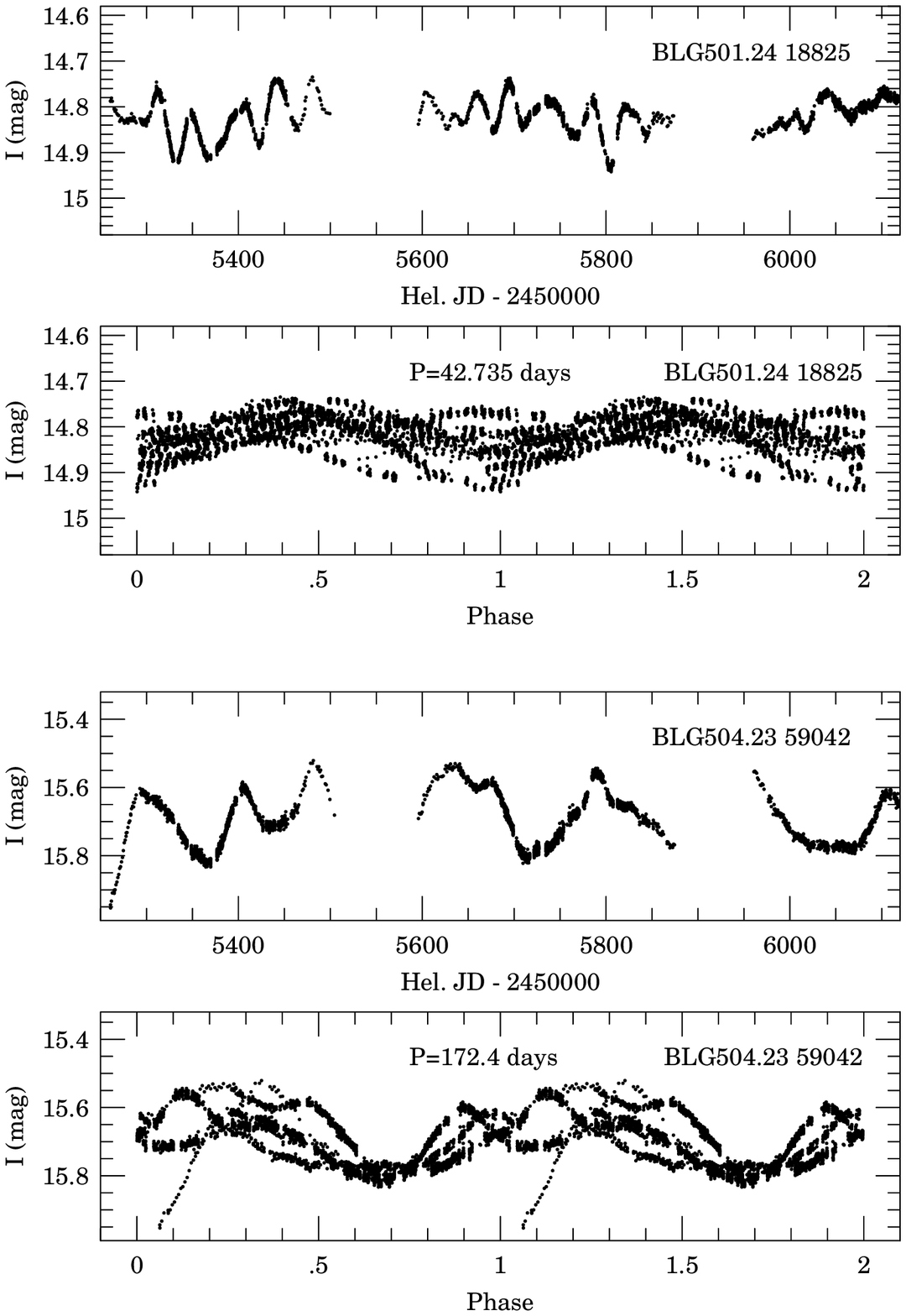}}
\FigCap{Light curves of pulsating red giants. Two {\it upper panels} show time
series and phased with the period 42.735 days light curve of the
OSARG-type optical counterpart candidate (BLG501.24 18825) to the GBS \#715
X-ray source. Two {\it bottom panels} show the light curves of the SRV
candidate (BLG504.23 59042; $P=172.4$~days) to the GBS \#907 X-ray
source.}
\end{figure}

Eclipsing stars subgroup includes 36 objects with clear eclipses. Contact
and detached short period binary systems are usually chromospherically
active due to fast rotation tidally synchronized with the short orbital
period and as a consequence -- strong X-ray sources. Upper panels of Fig.~4
show the sample light curves of stars from this class.
\begin{figure}[p]
\hglue-7mm{\includegraphics[width=13cm, bb=100 125 505 710]{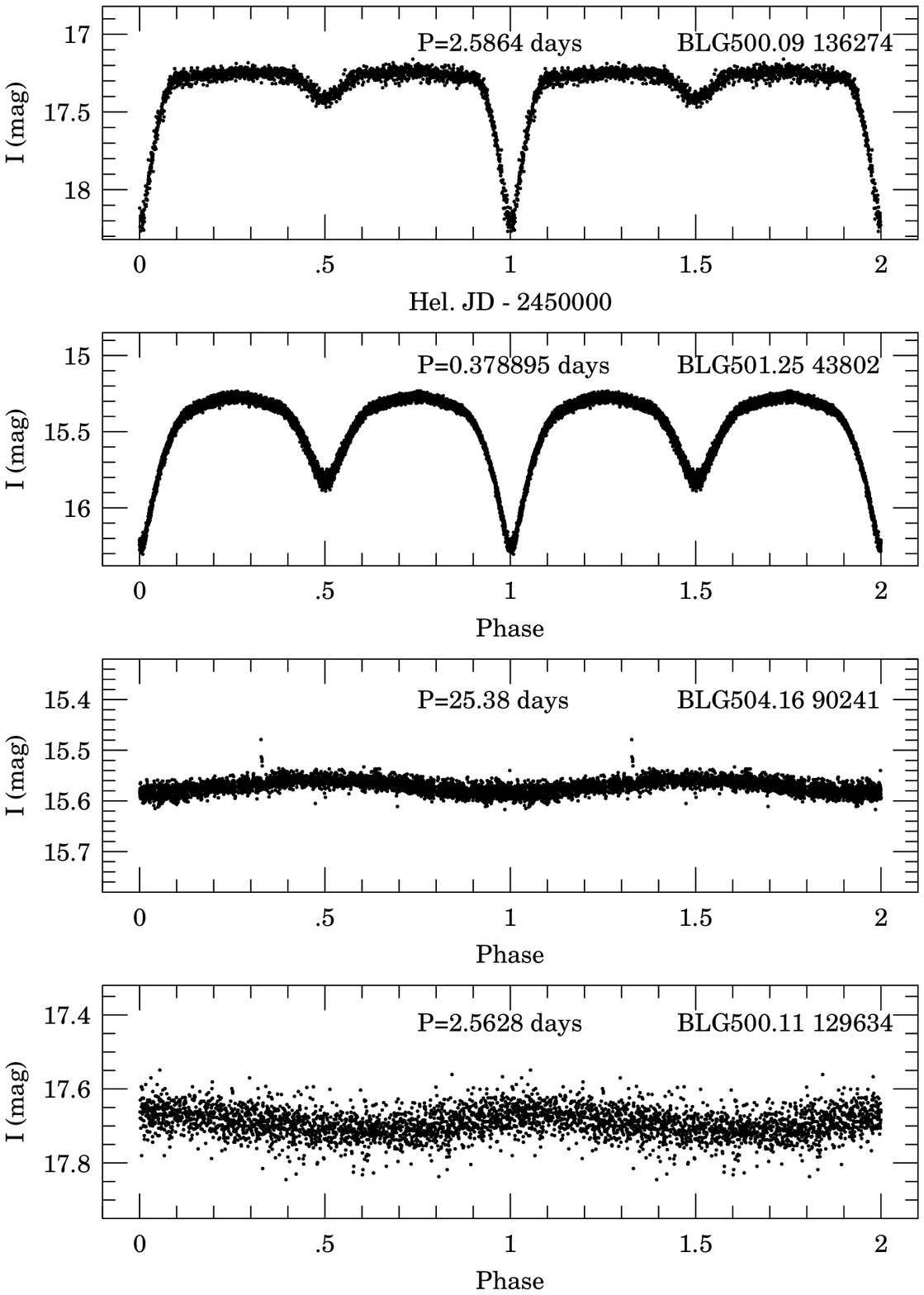}}
\FigCap{Phased light curves of eclipsing and periodic counterpart
candidates to the GBS X-ray sources: GBS \#684 (BLG500.09 136274), GBS
\#152 (BLG501.25 43802), GBS \#289 (BLG504.16 90241) and GBS \#428
(BLG500.11 129634).}
\end{figure}
\begin{figure}[htb]
\includegraphics[width=11.3cm, bb=100 280 505 710]{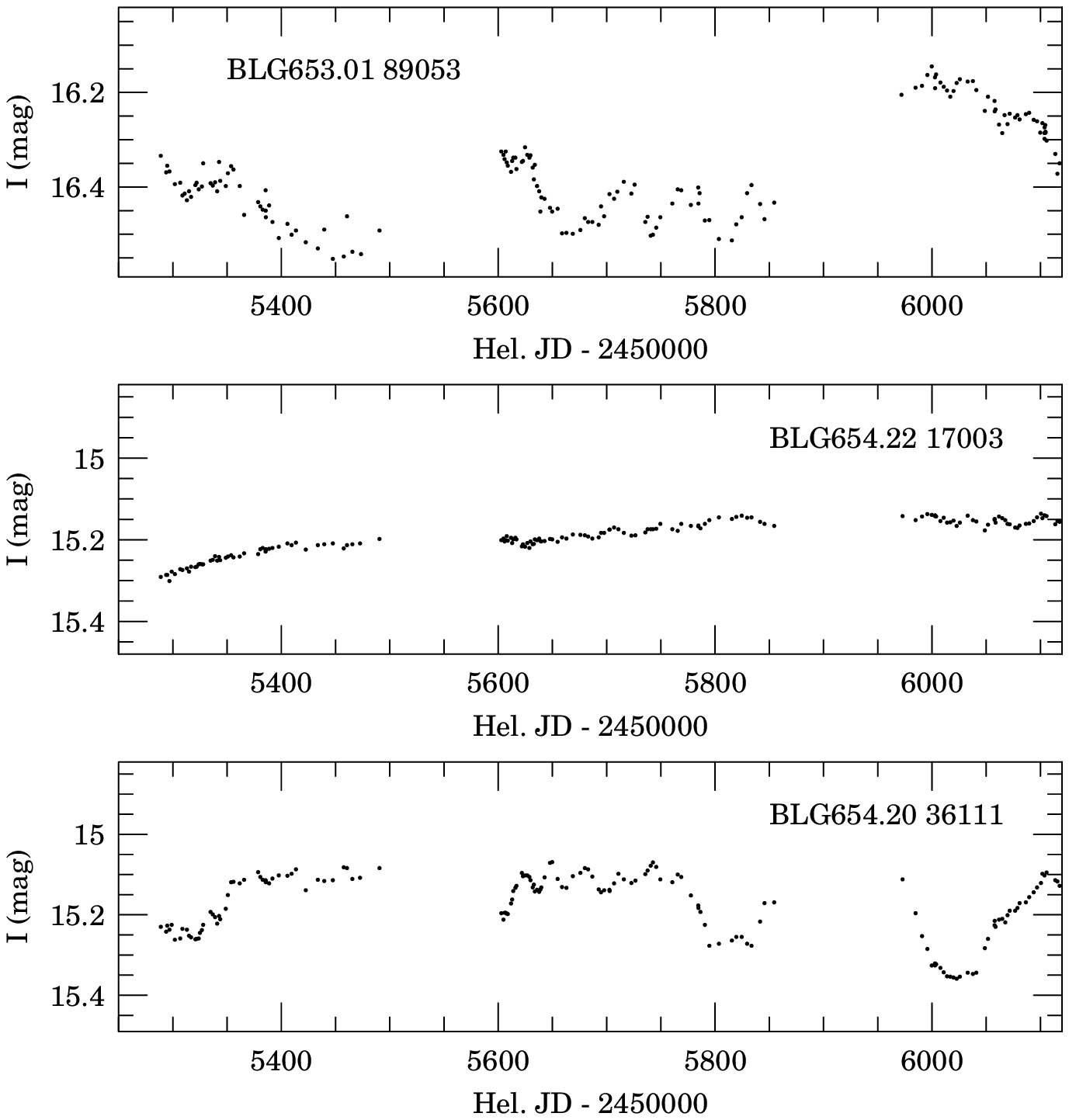}
\FigCap{Time series of the irregular counterpart candidates to the GBS
X-ray sources: GBS \#2 (BLG653.01 89053, Seyfert galaxy GRS 1734--292),
GBS \#212 (BLG654.22.17003) and GBS \#332 (BLG654.20.36111).}
\end{figure}
\begin{figure}[htb]
\includegraphics[width=11.5cm, bb=100 280 505 710]{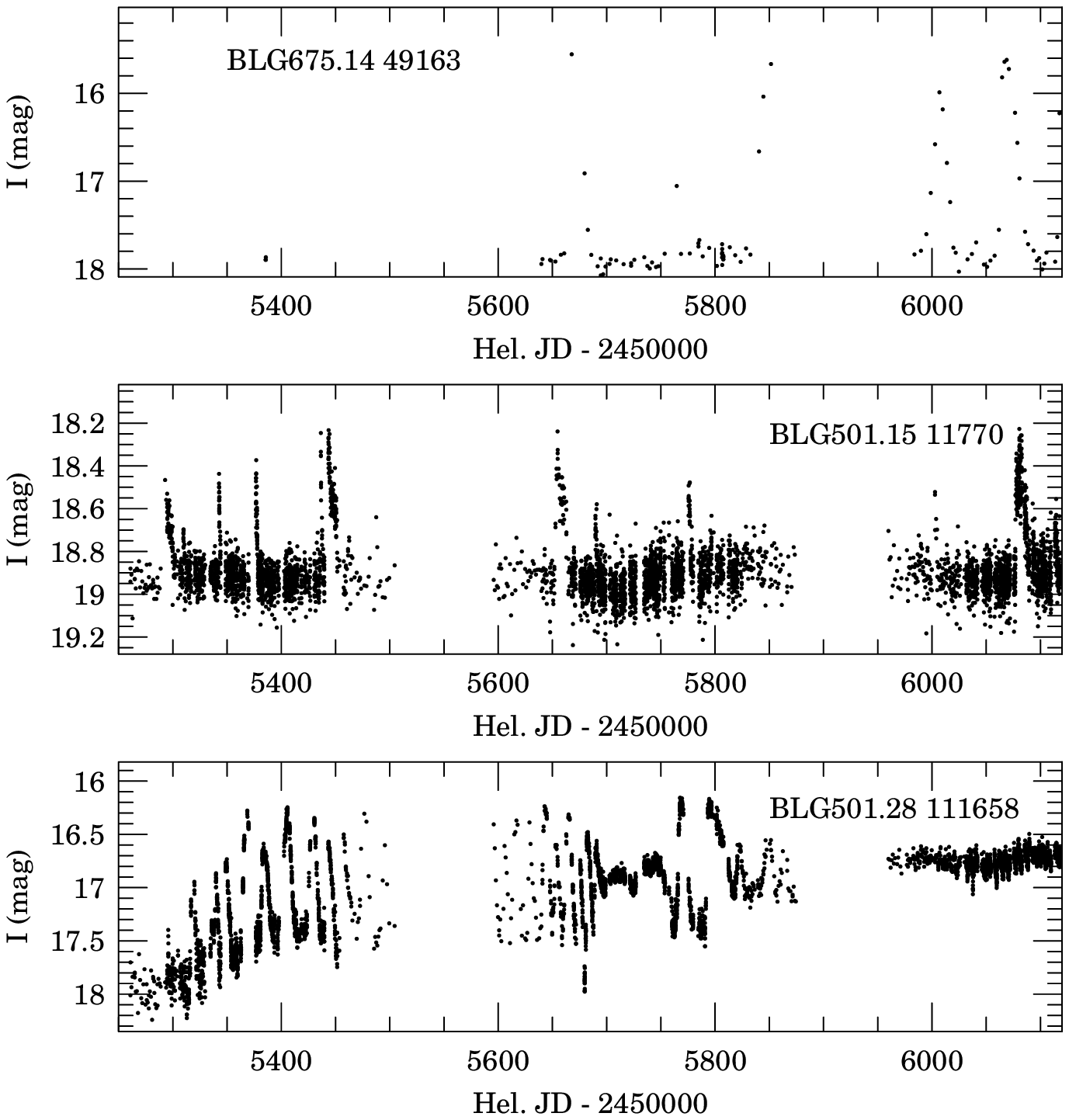}
\FigCap{Light curves of cataclysmic variables -- optical counterpart
candidates to the X-ray sources: GBS \#18 (BLG675.14 49163), GBS \#714
(BLG501.15 11770) and GBS \#426 (BLG501.28 111658).}
\end{figure}
\begin{figure}[htb]
\includegraphics[width=11.3cm, bb=100 280 505 710]{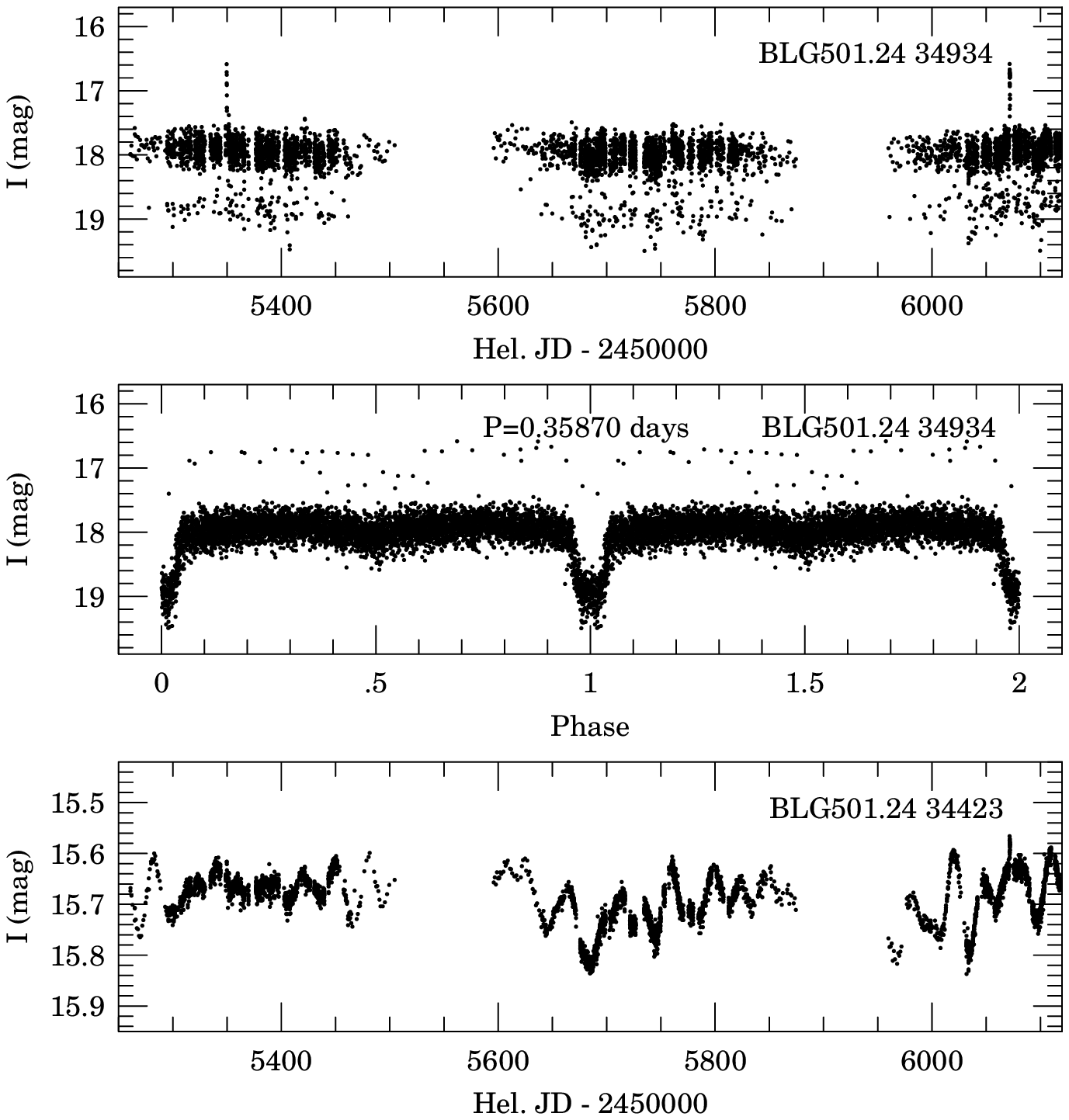}
\FigCap{Light curves of optical counterpart candidates to the X-ray
source GBS \#19. {\it Upper} and {\it middle panels:} eclipsing system
BLG501.24 34934. {\it Lower panel:} pulsating red giant BLG501.24 34423.}
\end{figure}

Two RR~Lyr pulsating stars, one of RRab and one of RRc type, were also
found in the selected sample of variable stars. Although pulsating stars
(Cepheids, RR~Lyr) can potentially be soft X-ray emitters only a handful
of Cepheids and RR Lyr stars have been detected in the X-ray band so far
(Engle and Guinan 2012). Our detection of the coincidence of the
positions of two RR~Lyr stars with X-ray sources, could be thus an
important discovery. However, we suspect that this may be a by chance
coincidence. The number of RR~Lyr stars detected in the typical Galactic
bulge field analyzed in this paper is about 40 per $2048\times4102$
pixel OGLE-IV CCD chip according to Soszyñski \etal (2011a) OGLE Catalog
of Variable Stars. Thus, we may expect about two RR~Lyr variable star
detections in the probed circle of 15 pixel radius in about 700 random
directions toward the Galactic center. Close examination of the finding
charts also suggest a by chance coincidence -- both RR~Lyr stars are
relatively far from the X-ray position, one almost at the boundary of
the searched circle.

The variability of the remaining periodic stars (48 objects) may be caused
either by the orbital motion (ellipsoidal variations), if they are
binaries, or by the rotation. In the case of ellipsoidal variation the real
period is twice of that found. Typical examples of stars from this group
are shown in the bottom panels of Fig.~4.

Irregular variables form a separate group. This kind of variability may
be an indicator of accreting systems like high or low mass X-ray
binaries or nova-like variables (flickering). It cannot be ruled out,
however, that some of the objects from this subgroup are actually
quasi-periodic but the small number of observations collected so far
prevented sound period detection. 

It is also worth noticing that the optical counterpart to the GBS \#2
source, which is a known Seyfert galaxy, GRS 1734--292 (Mart{\'\i} \etal
1998), is classified as irregular variable. OGLE provides, thus, long
term optical monitoring of this interesting AGN. Three examples of
irregular variables from this subgroup, including the counterpart of GBS
\#2, are shown in Fig.~5.

We also found three classical cataclysmic variables, two typical dwarf
novae and one Z~Cam type star. Cataclysmic variables are well known X-ray
emitters so the detection of these stars near the X-ray sources position
clearly indicates that they are optical counterparts to these X-ray
sources. Fig.~6 shows the light curves of these eruptive variables.

Table~1 lists all the X-ray sources from the GBS catalog cross-identified
with OGLE variable objects. For each optical counterpart candidate the
OGLE equatorial coordinates and distance from the X-ray source (in
arcsecs) are provided, its mean optical photometry ({\it I}-band magnitude,
$V-I$ color, if available), period in days in the case of periodic variable
stars and classification. Entries of counterpart candidates found in the
OGLE-II databases are marked with italic font. 

\renewcommand{\TableFont}{\scriptsize}
\MakeTableSep{r@{\hspace{7pt}}c@{\hspace{3pt}}r@{\hspace{7pt}}c@{\hspace{7pt}}c@{\hspace{7pt}}c@{\hspace{7pt}}c@{\hspace{7pt}}c@{\hspace{5pt}}r@{\hspace{0pt}}c@{\hspace{0pt}}l@{\hspace{5pt}}l}
{12.5cm}{Optical counterparts to the X-ray sources detected by the Galactic Bulge
Survey revealing optical variability}
{\hline
\noalign{\vskip3pt}
     GBS               & OGLE-IV & OGLE-IV              &    RA    &   DEC    & $D$     & $I$   & $V-I$ & \multicolumn{3}{c}{$P$}    & Remarks\\
\multicolumn{1}{c}{ID} & Field   &\multicolumn{1}{c}{No}& [2000.0] & [2000.0] & [\arcs] & [mag] & [mag] & \multicolumn{3}{c}{[days]} &\\   
\noalign{\vskip3pt}
\hline
\noalign{\vskip3pt}
         2 & BLG653.01 &  89053 & 17\uph37\upm28\zdot\ups38 & $-29\arcd08\arcm02\zdot\arcs1$ & 0.16 & 16.422 & --   &   --&&       & I AGN\\
        16 & BLG504.24 &     65 & 17\uph55\upm45\zdot\ups84 & $-27\arcd58\arcm13\zdot\arcs9$ & 0.18 & 14.572 & 2.383 &   4&.&95     & SP\\
        18 & BLG675.14 &  49163 & 17\uph39\upm35\zdot\ups76 & $-27\arcd29\arcm35\zdot\arcs9$ & 0.15 & 17.819 & --   &   --&&       & CV\\
 {\bf  19} & BLG501.24 &  34423 & 17\uph49\upm54\zdot\ups52 & $-29\arcd43\arcm35\zdot\arcs8$ & 0.66 & 15.668 & 5.938 &  31&.&65     & OSARG\\
 {\bf  19} & BLG501.24 &  34934 & 17\uph49\upm54\zdot\ups63 & $-29\arcd43\arcm35\zdot\arcs4$ & 0.87 & 17.983 & 1.519 &   0&.&3587   & E\\
        23 & BLG675.02 &  36509 & 17\uph42\upm31\zdot\ups57 & $-27\arcd43\arcm48\zdot\arcs3$ & 0.88 & 14.960 & --   &  44&.&63     & OSARG\\
        24 & BLG501.16 &  59134 & 17\uph48\upm49\zdot\ups50 & $-30\arcd01\arcm10\zdot\arcs3$ & 0.40 & 12.342 & 1.711 &  24&.&038    & SP\\
        29 & BLG500.27 &  22801 & 17\uph53\upm41\zdot\ups90 & $-28\arcd03\arcm53\zdot\arcs7$ & 0.37 & 15.619 & 2.134 &  12&.&771    & SP\\
        56 & BLG501.32 &   6608 & 17\uph50\upm00\zdot\ups36 & $-29\arcd24\arcm12\zdot\arcs7$ & 1.23 & 13.706 & 2.558 &  21&.&436    & E SP\\
        65 & BLG500.13 &  14897 & 17\uph51\upm35\zdot\ups26 & $-28\arcd43\arcm45\zdot\arcs3$ & 0.42 & 12.977 & 0.909 &   0&.&70676  & SP\\
        71 & BLG675.06 &   5524 & 17\uph39\upm49\zdot\ups29 & $-27\arcd54\arcm20\zdot\arcs6$ & 0.28 & 14.056 & --   &  --&&        & I\\
        89 & BLG653.11 &  44462 & 17\uph36\upm24\zdot\ups28 & $-28\arcd45\arcm32\zdot\arcs7$ & 0.24 & 14.228 & --   &  18&.&382    & SP\\
       103 & BLG501.29 &  50854 & 17\uph51\upm55\zdot\ups79 & $-29\arcd23\arcm10\zdot\arcs1$ & 3.72 & 14.714 & 1.684 &  --&&        & I\\
       106 & BLG500.08 & 191736 & 17\uph54\upm34\zdot\ups47 & $-28\arcd41\arcm48\zdot\arcs9$ & 0.37 & 12.908 & 1.676 &  32&.&57329  & SP\\
       122 & BLG653.08 &  10840 & 17\uph38\upm37\zdot\ups16 & $-28\arcd45\arcm44\zdot\arcs9$ & 0.44 & 13.369 & --   &  43&.&668    & SP\\
{\it   124}&{\it BUL\_SC5}& {\it 134677} & {\it 17\uph50\upm07\zdot\ups00} & ${\it -30\arcd01\arcm53\zdot\arcs9}$ & {\it 0.77} & {\it 12.481} & {\it 2.097} &  --&&  & {\it I}\\
       127 & BLG500.18 &  12184 & 17\uph54\upm21\zdot\ups19 & $-28\arcd30\arcm02\zdot\arcs2$ & 2.87 & 16.552 & 3.901 & 144&.&93     & P\\
       133 & BLG501.29 &  45257 & 17\uph52\upm01\zdot\ups41 & $-29\arcd25\arcm07\zdot\arcs9$ & 1.04 & 13.191 & 3.700 &  19&.&292    & OSARG\\
       134 & BLG648.28 &  31469 & 17\uph47\upm14\zdot\ups97 & $-26\arcd16\arcm49\zdot\arcs0$ & 0.05 & 13.711 & --   &  18&.&526    & OSARG\\
       137 & BLG504.14 & 157126 & 17\uph55\upm53\zdot\ups26 & $-28\arcd16\arcm33\zdot\arcs9$ & 0.69 & 15.113 & 1.315 &   0&.&43104  & E SP\\
       143 & BLG675.26 &  84886 & 17\uph42\upm49\zdot\ups73 & $-26\arcd48\arcm23\zdot\arcs0$ & 0.23 & 13.946 & --   &   1&.&72473  & P\\
       144 & BLG654.25 &  20335 & 17\uph33\upm00\zdot\ups32 & $-29\arcd36\arcm14\zdot\arcs9$ & 0.47 & 14.469 & --   &   2&.&47771  & E\\
       152 & BLG501.25 &  43802 & 17\uph48\upm56\zdot\ups72 & $-29\arcd38\arcm35\zdot\arcs9$ & 1.05 & 15.461 & 1.384 &   0&.&378895 & E\\
       155 & BLG504.07 &  23237 & 17\uph55\upm42\zdot\ups07 & $-28\arcd27\arcm39\zdot\arcs5$ & 0.69 & 16.515 & 2.057 &   0&.&53454  & E\\
       163 & BLG504.15 &  66246 & 17\uph55\upm32\zdot\ups43 & $-28\arcd04\arcm31\zdot\arcs4$ & 0.33 & 15.041 & 2.501 &   2&.&02896  & E\\
       174 & BLG654.27 &  54219 & 17\uph36\upm58\zdot\ups35 & $-29\arcd23\arcm51\zdot\arcs7$ & 1.33 & 12.848 & --   & 217&.&4      & P\\
       180 & BLG500.20 & 154077 & 17\uph52\upm20\zdot\ups78 & $-28\arcd23\arcm40\zdot\arcs6$ & 0.56 & 13.544 & --   &   2&.&5896   & E\\
       185 & BLG654.21 &  59612 & 17\uph35\upm27\zdot\ups56 & $-29\arcd45\arcm16\zdot\arcs6$ & 0.53 & 12.936 & --   &  44&.&25     & P\\
       190 & BLG653.12 &      5 & 17\uph35\upm44\zdot\ups39 & $-28\arcd54\arcm00\zdot\arcs4$ & 0.92 & 14.678 & --   & 225&.&2      & MIRA\\
       195 & BLG654.30 &  14819 & 17\uph35\upm07\zdot\ups54 & $-29\arcd19\arcm05\zdot\arcs1$ & 0.29 & 14.712 & --   &  28&.&5      & E\\
       196 & BLG500.07 &   1605 & 17\uph50\upm03\zdot\ups35 & $-29\arcd11\arcm50\zdot\arcs4$ & 1.28 & 17.089 & 1.854 &   1&.&9972   & SP\\
       200 & BLG504.16 &  58311 & 17\uph54\upm45\zdot\ups46 & $-28\arcd07\arcm43\zdot\arcs6$ & 2.79 & 18.350 & --   &  72&.&464    & P\\
       203 & BLG653.05 &  39870 & 17\uph34\upm55\zdot\ups60 & $-28\arcd58\arcm56\zdot\arcs9$ & 2.11 & 13.949 & --   &   9&.&38086  & SP\\
       204 & BLG500.10 & 172048 & 17\uph53\upm12\zdot\ups45 & $-28\arcd47\arcm46\zdot\arcs5$ & 0.97 & 15.267 & 1.692 &   2&.&014    & SP\\
       212 & BLG654.22 &  17003 & 17\uph35\upm10\zdot\ups15 & $-29\arcd36\arcm06\zdot\arcs1$ & 0.78 & 15.209 & --   &  --&&        & I\\
       216 & BLG653.09 &  79513 & 17\uph37\upm21\zdot\ups71 & $-28\arcd53\arcm57\zdot\arcs3$ & 0.32 & 13.216 & --   &  16&.&42036  & SP\\
       222 & BLG648.16 &   1012 & 17\uph44\upm05\zdot\ups45 & $-27\arcd00\arcm13\zdot\arcs3$ & 0.11 & 15.514 & --   &  46&.&083    & SP\\
       227 & BLG501.28 & 155033 & 17\uph52\upm19\zdot\ups58 & $-29\arcd21\arcm09\zdot\arcs8$ & 0.08 & 14.191 & 2.300 &   2&.&947    & SP\\
       228 & BLG534.32 &  97586 & 17\uph49\upm24\zdot\ups63 & $-30\arcd34\arcm44\zdot\arcs7$ & 0.42 & 13.089 & 1.835 &  29&.&916    & SP\\
       231 & BLG648.28 &  36747 & 17\uph47\upm14\zdot\ups73 & $-26\arcd09\arcm38\zdot\arcs2$ & 0.25 & 13.140 & --   &  69&.&32     & OSARG\\
       234 & BLG675.28 &  21714 & 17\uph41\upm47\zdot\ups68 & $-26\arcd46\arcm49\zdot\arcs8$ & 2.86 & 16.539 & --   &  26&.&455    & P\\
       237 & BLG654.17 &  56669 & 17\uph38\upm15\zdot\ups98 & $-29\arcd39\arcm39\zdot\arcs9$ & 0.72 & 14.138 & --   &   0&.&273483 & E\\
       240 & BLG654.21 &   7912 & 17\uph35\upm47\zdot\ups78 & $-29\arcd43\arcm33\zdot\arcs2$ & 0.56 & 12.523 & --   &  --&&        & I\\
       253 & BLG504.25 &  87539 & 17\uph54\upm27\zdot\ups60 & $-27\arcd47\arcm32\zdot\arcs9$ & 0.89 & 12.689 & 0.943 &   1&.&31734  & E SP\\
       282 & BLG653.11 &    550 & 17\uph36\upm34\zdot\ups43 & $-28\arcd53\arcm33\zdot\arcs9$ & 0.05 & 18.322 & --   &   9&.&72760  & P\\
 {\bf 289} & BLG504.16 &  90151 & 17\uph54\upm39\zdot\ups15 & $-28\arcd08\arcm50\zdot\arcs0$ & 0.32 & 14.723 & 2.400 &  11&.&84     & SP\\
 {\bf 289} & BLG504.16 &  90241 & 17\uph54\upm38\zdot\ups88 & $-28\arcd08\arcm51\zdot\arcs4$ & 3.56 & 15.575 & 4.250 &  25&.&38     & P\\
       295 & BLG500.10 & 186047 & 17\uph53\upm11\zdot\ups01 & $-28\arcd42\arcm27\zdot\arcs7$ & 0.25 & 13.432 & 1.689 &   2&.&92227  & SP\\
\noalign{\vskip3pt}
\hline}
\setcounter{table}{0}
\MakeTableSepp{r@{\hspace{7pt}}c@{\hspace{3pt}}r@{\hspace{7pt}}c@{\hspace{7pt}}c@{\hspace{7pt}}c@{\hspace{7pt}}c@{\hspace{7pt}}c@{\hspace{5pt}}r@{\hspace{0pt}}c@{\hspace{0pt}}l@{\hspace{5pt}}l}
{12.5cm}{Continued}
{\hline
\noalign{\vskip3pt}
     GBS               & OGLE-IV & OGLE-IV              &    RA    &   DEC    & $D$     & $I$   & $V-I$ & \multicolumn{3}{c}{$P$}    & Remarks\\
\multicolumn{1}{c}{ID} & Field   &\multicolumn{1}{c}{No}& [2000.0] & [2000.0] & [\arcs] & [mag] & [mag] & \multicolumn{3}{c}{[days]} &\\   
\noalign{\vskip3pt}
\hline
\noalign{\vskip3pt}
       300 & BLG654.14 &  66787 & 17\uph33\upm51\zdot\ups57 & $-30\arcd06\arcm30\zdot\arcs9$ & 0.62 & 15.282 & --   &  21&.&1417   & SP\\
       303 & BLG504.31 &  31226 & 17\uph56\upm02\zdot\ups11 & $-27\arcd39\arcm21\zdot\arcs2$ & 2.17 & 12.753 & 1.213 &   0&.&8313   & SP\\
       304 & BLG504.31 &  31226 & 17\uph56\upm02\zdot\ups11 & $-27\arcd39\arcm21\zdot\arcs2$ & 1.57 & 12.753 & 1.213 &   0&.&8313   & SP\\
       326 & BLG675.20 &  61377 & 17\uph41\upm31\zdot\ups36 & $-27\arcd09\arcm59\zdot\arcs2$ & 0.32 & 13.827 & --   &   0&.&23462  & P\\
       330 & BLG653.19 &  81200 & 17\uph36\upm43\zdot\ups88 & $-28\arcd21\arcm21\zdot\arcs3$ & 0.82 & 15.069 & --   &  --&&        & I\\
       331 & BLG654.28 &  28452 & 17\uph36\upm23\zdot\ups47 & $-29\arcd22\arcm34\zdot\arcs7$ & 0.14 & 17.277 & --   &  18&.&215    & SP\\
       332 & BLG654.20 &  36111 & 17\uph36\upm20\zdot\ups20 & $-29\arcd33\arcm38\zdot\arcs9$ & 0.80 & 15.156 & --   &  --&&        & I\\
       336 & BLG654.04 &  45052 & 17\uph35\upm31\zdot\ups42 & $-30\arcd24\arcm23\zdot\arcs3$ & 0.38 & 15.696 & --   &   9&.&99     & SP\\
       354 & BLG500.18 & 172616 & 17\uph53\upm52\zdot\ups75 & $-28\arcd25\arcm13\zdot\arcs2$ & 0.79 & 15.092 & 2.074 &   8&.&21018  & SP\\
{\it   362}& {\it BUL\_SC5}&{\it 284655} & {\it 17\uph50\upm37\zdot\ups53} & ${\it -29\arcd40\arcm45\zdot\arcs5}$ & {\it 0.26} & {\it 11.420} & {\it 2.272} & {\it 107}&.&{\it 53} & {\it P}\\
       363 & BLG501.31 &  34121 & 17\uph50\upm33\zdot\ups73 & $-29\arcd21\arcm33\zdot\arcs5$ & 0.35 & 14.005 & 1.679 &   8&.&7413   & SP\\
       375 & BLG675.17 &  11128 & 17\uph43\upm57\zdot\ups95 & $-27\arcd10\arcm34\zdot\arcs0$ & 1.00 & 15.365 & --   &  62&.&8      & SP\\
       387 & BLG654.27 &  27852 & 17\uph37\upm05\zdot\ups58 & $-29\arcd24\arcm52\zdot\arcs5$ & 0.86 & 13.044 & --   &   4&.&831    & P\\
       391 & BLG654.11 &  47010 & 17\uph36\upm06\zdot\ups42 & $-30\arcd03\arcm11\zdot\arcs0$ & 0.33 & 13.977 & --   &  11&.&47     & E SP\\
       395 & BLG654.31 &  30300 & 17\uph34\upm18\zdot\ups96 & $-29\arcd28\arcm33\zdot\arcs8$ & 1.12 & 14.868 & --   &   5&.&965    & SP\\
       397 & BLG654.23 &  61348 & 17\uph33\upm59\zdot\ups54 & $-29\arcd49\arcm47\zdot\arcs2$ & 0.50 & 13.190 & --   &   0&.&77471  & E\\
       414 & BLG504.30 &   6878 & 17\uph57\upm09\zdot\ups75 & $-27\arcd32\arcm49\zdot\arcs7$ & 0.25 & 13.715 & 1.585 &   0&.&40826  & SP\\
       416 & BLG504.16 &  83867 & 17\uph54\upm35\zdot\ups04 & $-28\arcd11\arcm52\zdot\arcs5$ & 1.43 & 14.998 & --   &  16&.&5563   & P\\
       417 & BLG504.16 & 130997 & 17\uph54\upm27\zdot\ups75 & $-28\arcd07\arcm49\zdot\arcs3$ & 0.29 & 13.749 & 0.532 &   0&.&95923  & P\\
       426 & BLG501.28 & 111658 & 17\uph52\upm36\zdot\ups05 & $-29\arcd19\arcm39\zdot\arcs9$ & 0.17 & 17.293 & 0.439 &  --&&        & CV\\
       428 & BLG500.11 & 129634 & 17\uph52\upm32\zdot\ups88 & $-28\arcd37\arcm48\zdot\arcs5$ & 0.55 & 17.698 & 1.710 &   2&.&56279  & P\\
       429 & BLG500.20 & 120943 & 17\uph52\upm32\zdot\ups43 & $-28\arcd21\arcm01\zdot\arcs1$ & 0.28 & 16.839 & 2.037 &  10&.&94092  & SP\\
       432 & BLG500.13 &  29005 & 17\uph51\upm19\zdot\ups17 & $-28\arcd53\arcm08\zdot\arcs9$ & 0.49 & 14.163 & 1.781 &  11&.&876    & SP\\
       434 & BLG500.05 &  84278 & 17\uph51\upm10\zdot\ups63 & $-29\arcd01\arcm17\zdot\arcs7$ & 0.62 & 14.479 & 1.101 &   0&.&69871  & SP\\
       439 & BLG534.25 &  65908 & 17\uph48\upm40\zdot\ups65 & $-30\arcd55\arcm53\zdot\arcs2$ & 1.75 & 16.443 & --   &  64&.&516    & SRV\\
       472 & BLG675.22 &  60741 & 17\uph40\upm25\zdot\ups65 & $-27\arcd05\arcm40\zdot\arcs4$ & 0.29 & 14.623 & --   &  21&.&834    & SP\\
       483 & BLG654.26 &  59596 & 17\uph37\upm31\zdot\ups10 & $-29\arcd24\arcm28\zdot\arcs8$ & 3.47 & 13.476 & --   &   0&.&6462   & E\\
       484 & BLG654.19 &  17980 & 17\uph37\upm19\zdot\ups57 & $-29\arcd33\arcm23\zdot\arcs8$ & 0.57 & 16.433 & --   &   7&.&062    & SP\\
       491 & BLG653.04 &  15499 & 17\uph35\upm43\zdot\ups46 & $-29\arcd03\arcm11\zdot\arcs8$ & 1.99 & 14.561 & --   &  16&.&10306  & P\\
       494 & BLG654.05 &  47692 & 17\uph34\upm58\zdot\ups69 & $-30\arcd13\arcm29\zdot\arcs0$ & 1.74 & 16.667 & --   &  --&&        & I\\
       495 & BLG654.22 &  76376 & 17\uph34\upm47\zdot\ups61 & $-29\arcd35\arcm10\zdot\arcs5$ & 2.09 & 16.623 & --   &   0&.&48629  & P\\
       497 & BLG654.14 &   1240 & 17\uph34\upm22\zdot\ups10 & $-30\arcd05\arcm05\zdot\arcs9$ & 0.20 & 14.057 & --   &   3&.&4965   & SP\\
       500 & BLG654.16 &  68256 & 17\uph32\upm39\zdot\ups40 & $-30\arcd02\arcm22\zdot\arcs7$ & 0.22 & 15.163 & --   &   1&.&8954   & P\\
       501 & BLG500.11 &  55525 & 17\uph52\upm52\zdot\ups63 & $-28\arcd48\arcm05\zdot\arcs4$ & 0.67 & 14.142 & 1.975 &  15&.&03759  & SP\\
       502 & BLG500.20 & 154130 & 17\uph52\upm23\zdot\ups57 & $-28\arcd24\arcm10\zdot\arcs9$ & 1.38 & 15.218 & 1.471 &   6&.&9252   & SP\\
       505 & BLG648.16 &  31001 & 17\uph43\upm53\zdot\ups28 & $-26\arcd53\arcm53\zdot\arcs7$ & 0.67 & 13.048 & --   &  11&.&147    & SP\\
       513 & BLG504.22 & 107805 & 17\uph56\upm46\zdot\ups04 & $-27\arcd45\arcm47\zdot\arcs6$ & 0.25 & 13.097 & 2.450 &  11&.&8343   & OSARG\\
       515 & BLG504.24 &     65 & 17\uph55\upm45\zdot\ups84 & $-27\arcd58\arcm13\zdot\arcs9$ & 3.14 & 14.572 & 2.383 &   4&.&95     & SP\\
       516 & BLG504.24 &  57283 & 17\uph55\upm34\zdot\ups69 & $-27\arcd47\arcm59\zdot\arcs7$ & 3.12 & 15.751 & --   &  33&.&333    & SRV\\
       522 & BLG500.17 & 158208 & 17\uph54\upm32\zdot\ups61 & $-28\arcd29\arcm19\zdot\arcs6$ & 1.96 & 12.971 & 5.007 & 147&.&3      & SRV\\
       526 & BLG500.01 & 144233 & 17\uph54\upm01\zdot\ups04 & $-29\arcd00\arcm46\zdot\arcs2$ & 3.71 & 15.429 & 2.513 & 135&.&1      & P\\
       532 & BLG500.21 &   2338 & 17\uph52\upm15\zdot\ups84 & $-28\arcd34\arcm15\zdot\arcs3$ & 0.73 & 12.890 & 1.873 &  44&.&25     & SP\\
       538 & BLG501.22 & 116664 & 17\uph50\upm54\zdot\ups10 & $-29\arcd46\arcm16\zdot\arcs0$ & 0.60 & 13.799 & 1.176 &   6&.&0533   & SP\\
       543 & BLG501.24 &  78370 & 17\uph49\upm32\zdot\ups70 & $-29\arcd49\arcm55\zdot\arcs9$ & 0.41 & 14.280 & 2.918 &  61&.&35     & SP\\
       556 & BLG648.29 &  41317 & 17\uph46\upm30\zdot\ups18 & $-26\arcd22\arcm04\zdot\arcs4$ & 0.20 & 14.675 & --   &  --&&        & I\\
       561 & BLG648.30 &   8035 & 17\uph46\upm07\zdot\ups68 & $-26\arcd15\arcm47\zdot\arcs1$ & 0.54 & 16.121 & --   &  12&.&4915   & E\\
       575 & BLG648.14 &  72850 & 17\uph44\upm59\zdot\ups28 & $-26\arcd51\arcm10\zdot\arcs2$ & 0.68 & 14.951 & --   &   6&.&5147   & SP\\
       594 & BLG675.18 &  76442 & 17\uph42\upm46\zdot\ups29 & $-27\arcd13\arcm53\zdot\arcs3$ & 0.16 & 16.472 & --   &  27&.&322    & SP\\
 {\bf 611} & BLG675.23 &  21477 & 17\uph39\upm50\zdot\ups50 & $-27\arcd09\arcm16\zdot\arcs5$ & 1.70 & 13.201 & --   &   1&.&3776   & P\\
 {\bf 611} & BLG675.23 &  21504 & 17\uph39\upm50\zdot\ups34 & $-27\arcd09\arcm17\zdot\arcs3$ & 2.98 & 14.793 & --   &  --&&        & I\\
       633 & BLG654.15 &  41782 & 17\uph33\upm18\zdot\ups72 & $-30\arcd05\arcm22\zdot\arcs3$ & 0.67 & 14.615 & --   &   0&.&73185  & P\\
       634 & BLG654.24 &  67281 & 17\uph33\upm16\zdot\ups65 & $-29\arcd40\arcm03\zdot\arcs2$ & 0.20 & 14.744 & --   &   0&.&76104  & SP\\
\noalign{\vskip3pt}
\hline}
\setcounter{table}{0}
\MakeTableSepp{r@{\hspace{7pt}}c@{\hspace{3pt}}r@{\hspace{7pt}}c@{\hspace{7pt}}c@{\hspace{7pt}}c@{\hspace{7pt}}c@{\hspace{7pt}}c@{\hspace{5pt}}r@{\hspace{0pt}}c@{\hspace{0pt}}l@{\hspace{5pt}}l}
{12.5cm}{Continued}
{\hline
\noalign{\vskip3pt}
     GBS               & OGLE-IV & OGLE-IV              &    RA    &   DEC    & $D$     & $I$   & $V-I$ & \multicolumn{3}{c}{$P$}    & Remarks\\
\multicolumn{1}{c}{ID} & Field   &\multicolumn{1}{c}{No}& [2000.0] & [2000.0] & [\arcs] & [mag] & [mag] & \multicolumn{3}{c}{[days]} &\\   
\noalign{\vskip3pt}
\hline
\noalign{\vskip3pt}
       637 & BLG654.15 &  75839 & 17\uph33\upm08\zdot\ups55 & $-29\arcd57\arcm27\zdot\arcs6$ & 1.21 & 13.429 & --   &   4&.&97018  & P\\
       642 & BLG504.14 & 192340 & 17\uph55\upm55\zdot\ups14 & $-28\arcd01\arcm58\zdot\arcs9$ & 1.12 & 14.924 & 1.948 &   2&.&924    & SP\\
       649 & BLG504.15 &  98233 & 17\uph55\upm24\zdot\ups65 & $-28\arcd10\arcm36\zdot\arcs8$ & 0.79 & 13.431 & 2.935 &  30&.&67485  & SP\\
       650 & BLG500.09 &  64937 & 17\uph54\upm08\zdot\ups51 & $-28\arcd48\arcm28\zdot\arcs8$ & 3.29 & 12.849 & 4.751 &  31&.&5      & SRV\\
       651 & BLG648.28 &  48495 & 17\uph47\upm03\zdot\ups70 & $-26\arcd11\arcm08\zdot\arcs8$ & 0.18 & 13.485 & --   &  --&&        & I\\
       657 & BLG504.29 &  15566 & 17\uph57\upm46\zdot\ups88 & $-27\arcd33\arcm09\zdot\arcs8$ & 0.39 & 16.280 & 1.562 &  16&.&58375  & P\\
       659 & BLG645.04 &  31747 & 17\uph57\upm38\zdot\ups37 & $-27\arcd19\arcm55\zdot\arcs2$ & 0.28 & 14.556 & --   &  74&.&627    & SP\\
       663 & BLG504.30 &  69448 & 17\uph56\upm30\zdot\ups96 & $-27\arcd31\arcm39\zdot\arcs9$ & 0.53 & 13.495 & 2.450 &  24&.&691    & SP\\
       666 & BLG504.23 & 118900 & 17\uph56\upm01\zdot\ups69 & $-27\arcd44\arcm48\zdot\arcs8$ & 0.49 & 14.104 & 1.522 &   2&.&6201   & E SP\\
       668 & BLG504.14 & 188107 & 17\uph55\upm55\zdot\ups86 & $-28\arcd05\arcm55\zdot\arcs5$ & 0.41 & 14.147 & 2.669 & 106&.&4      & SP\\
       674 & BLG504.24 &  74705 & 17\uph55\upm19\zdot\ups78 & $-27\arcd56\arcm32\zdot\arcs5$ & 1.07 & 14.868 & --   &  24&.&73     & OSARG\\
       676 & BLG500.17 & 159235 & 17\uph54\upm32\zdot\ups45 & $-28\arcd28\arcm13\zdot\arcs4$ & 3.73 & 17.241 & 2.748 &   0&.&30728  & RRLYR\\
       679 & BLG500.18 &  50697 & 17\uph54\upm13\zdot\ups98 & $-28\arcd33\arcm41\zdot\arcs6$ & 3.68 & 15.111 & 3.425 &  --&&        & I\\
       681 & BLG500.09 & 128064 & 17\uph54\upm05\zdot\ups63 & $-28\arcd47\arcm32\zdot\arcs6$ & 0.04 & 13.865 & 0.826 &   2&.&0736   & SP\\
       684 & BLG500.09 & 136274 & 17\uph54\upm00\zdot\ups38 & $-28\arcd43\arcm32\zdot\arcs4$ & 2.47 & 17.369 & 2.566 &   2&.&5864   & E SP\\
       685 & BLG500.09 & 128146 & 17\uph53\upm56\zdot\ups51 & $-28\arcd46\arcm58\zdot\arcs3$ & 0.58 & 14.760 & 1.345 &   2&.&7894   & SP\\
       688 & BLG500.19 &    348 & 17\uph53\upm42\zdot\ups30 & $-28\arcd35\arcm31\zdot\arcs8$ & 0.63 & 16.640 & 3.175 &   8&.&231    & SP\\
       689 & BLG500.19 &  17579 & 17\uph53\upm39\zdot\ups84 & $-28\arcd29\arcm31\zdot\arcs6$ & 0.36 & 13.825 & 1.525 & 189& &       & P\\
       691 & BLG500.19 & 125229 & 17\uph53\upm24\zdot\ups21 & $-28\arcd27\arcm10\zdot\arcs8$ & 1.20 & 19.629 & --   &   2&.&794    & P\\
       692 & BLG500.02 & 163813 & 17\uph53\upm05\zdot\ups18 & $-29\arcd13\arcm00\zdot\arcs6$ & 1.33 & 14.609 & 3.796 &  93&.&3      & P\\
       695 & BLG501.28 &  64217 & 17\uph52\upm40\zdot\ups19 & $-29\arcd20\arcm57\zdot\arcs8$ & 0.26 & 15.703 & 2.757 &   0&.&30988  & P\\
       701 & BLG501.29 & 138642 & 17\uph51\upm41\zdot\ups10 & $-29\arcd18\arcm55\zdot\arcs2$ & 0.77 & 13.531 & 1.838 &  11&.&9936   & E SP\\
       702 & BLG500.12 & 122438 & 17\uph51\upm39\zdot\ups41 & $-28\arcd52\arcm43\zdot\arcs5$ & 0.44 & 15.540 & 1.560 &   1&.&50011  & E\\
 {\bf 705} & BLG501.30 &  35398 & 17\uph51\upm16\zdot\ups15 & $-29\arcd23\arcm44\zdot\arcs4$ & 0.87 & 15.975 & 1.579 &  --&&        & I \\
 {\bf 705} & BLG501.30 &  35400 & 17\uph51\upm15\zdot\ups91 & $-29\arcd23\arcm42\zdot\arcs8$ & 2.76 & 15.400 & --   & 326&.&8      & MIRA\\
       711 & BLG501.31 &  34198 & 17\uph50\upm32\zdot\ups49 & $-29\arcd21\arcm00\zdot\arcs1$ & 0.68 & 15.649 & 2.322 &  13&.&73     & SP\\
       712 & BLG501.14 & 108924 & 17\uph50\upm14\zdot\ups77 & $-30\arcd03\arcm19\zdot\arcs5$ & 0.76 & 15.556 & 2.485 &   6&.&84     & SP\\
       714 & BLG501.15 &  11770 & 17\uph49\upm59\zdot\ups31 & $-29\arcd57\arcm11\zdot\arcs1$ & 0.71 & 18.874 & --   &  --&&        & CV\\
       715 & BLG501.24 &  18825 & 17\uph49\upm59\zdot\ups15 & $-29\arcd34\arcm38\zdot\arcs1$ & 0.79 & 14.836 & --   &  42&.&735    & OSARG\\
       717 & BLG501.24 &  35909 & 17\uph49\upm50\zdot\ups32 & $-29\arcd42\arcm30\zdot\arcs4$ & 2.92 & 18.600 & --   &   0&.&54767  & RRLYR$^*$\\
       718 & BLG501.07 &  45474 & 17\uph49\upm48\zdot\ups54 & $-30\arcd12\arcm21\zdot\arcs0$ & 0.65 & 15.335 & 1.437 &   5&.&824    & P\\
       721 & BLG534.32 & 107583 & 17\uph49\upm20\zdot\ups74 & $-30\arcd28\arcm05\zdot\arcs9$ & 0.34 & 15.633 & --   &   2&.&66207  & E\\
{\it   724}& {\it BUL\_SC44} & {\it 12986} & {\it 17\uph48\upm54\zdot\ups26} & ${\it -30\arcd18\arcm39\zdot\arcs5}$ & {\it 0.51} & {\it 12.827} & {\it 1.104} &   {\it 0}&.&{\it 32200}  & {\it E}\\
       742 & BLG648.29 &  69672 & 17\uph46\upm19\zdot\ups94 & $-26\arcd12\arcm28\zdot\arcs7$ & 0.11 & 14.430 & --   &   0&.&25321  & P\\
       774 & BLG675.17 &   4519 & 17\uph44\upm01\zdot\ups34 & $-27\arcd16\arcm46\zdot\arcs8$ & 3.57 & 14.530 & --   &  43&.&47826  & P\\
       778 & BLG648.16 &  78697 & 17\uph43\upm35\zdot\ups43 & $-26\arcd48\arcm57\zdot\arcs1$ & 0.44 & 17.822 & --   &   6&.&079    & P\\
       794 & BLG675.03 &  22881 & 17\uph41\upm42\zdot\ups38 & $-27\arcd58\arcm30\zdot\arcs3$ & 0.78 & 19.209 & --   &   0&.&11786  & E?\\
       796 & BLG675.03 &  73380 & 17\uph41\upm27\zdot\ups62 & $-27\arcd58\arcm39\zdot\arcs2$ & 0.48 & 14.229 & --   &   1&.&06571  & E\\
       800 & BLG675.04 &  85070 & 17\uph40\upm51\zdot\ups59 & $-27\arcd45\arcm59\zdot\arcs7$ & 0.32 & 13.954 & --   & 213& &       & MIRA\\
       809 & BLG675.22 &  60737 & 17\uph40\upm17\zdot\ups79 & $-27\arcd06\arcm07\zdot\arcs8$ & 1.26 & 14.581 & --   &  35&.&7      & P\\
       811 & BLG675.14 &  28661 & 17\uph39\upm48\zdot\ups50 & $-27\arcd24\arcm16\zdot\arcs0$ & 0.59 & 14.668 & --   &  43&.&3      & P\\
       819 & BLG675.07 &  56138 & 17\uph38\upm56\zdot\ups46 & $-27\arcd48\arcm33\zdot\arcs4$ & 0.36 & 16.978 & --   &  35& &       & SP\\
       829 & BLG654.26 &  41915 & 17\uph37\upm38\zdot\ups06 & $-29\arcd21\arcm44\zdot\arcs8$ & 0.42 & 15.767 & --   &  --&&        & I\\
       830 & BLG653.26 &  52494 & 17\uph37\upm29\zdot\ups94 & $-28\arcd07\arcm59\zdot\arcs5$ & 0.58 & 13.127 & --   &   8&.&288    & E\\
       837 & BLG653.19 &  61176 & 17\uph36\upm47\zdot\ups64 & $-28\arcd22\arcm21\zdot\arcs7$ & 2.41 & 15.714 & --   &  15&.&18     & P\\
       843 & BLG654.03 &  59545 & 17\uph36\upm05\zdot\ups70 & $-30\arcd18\arcm16\zdot\arcs8$ & 3.27 & 14.403 & --   &  49&.&5      & OSARG\\
       845 & BLG654.29 &   2771 & 17\uph35\upm47\zdot\ups48 & $-29\arcd29\arcm15\zdot\arcs0$ & 0.08 & 16.443 & --   &   3&.&158    & P\\
       847 & BLG654.12 &  30313 & 17\uph35\upm37\zdot\ups78 & $-29\arcd56\arcm01\zdot\arcs3$ & 1.00 & 14.470 & --   &  17&.&212    & P\\
       848 & BLG654.04 &  55881 & 17\uph35\upm33\zdot\ups43 & $-30\arcd17\arcm44\zdot\arcs0$ & 0.42 & 18.294 & --   &   7&.&4405   & SP\\
       852 & BLG654.13 &  32681 & 17\uph35\upm02\zdot\ups21 & $-30\arcd03\arcm54\zdot\arcs7$ & 0.93 & 14.742 & --   &   0&.&29857  & P\\
       853 & BLG654.13 &  74975 & 17\uph34\upm46\zdot\ups94 & $-29\arcd56\arcm31\zdot\arcs6$ & 0.64 & 16.118 & --   &   0&.&91405  & E\\
\noalign{\vskip3pt}
\hline
\noalign{\vskip3pt}
\multicolumn{12}{l}{$^*$ OGLE-BLG-RRLYR-03686 (Soszyñski \etal 2011a)}
}
\setcounter{table}{0}
\MakeTableSepp{r@{\hspace{7pt}}c@{\hspace{3pt}}r@{\hspace{7pt}}c@{\hspace{7pt}}c@{\hspace{7pt}}c@{\hspace{7pt}}c@{\hspace{7pt}}c@{\hspace{5pt}}r@{\hspace{0pt}}c@{\hspace{0pt}}l@{\hspace{5pt}}l}
{12.5cm}{Concluded}
{\hline
\noalign{\vskip3pt}
     GBS               & OGLE-IV & OGLE-IV              &    RA    &   DEC    & $D$     & $I$   & $V-I$ & \multicolumn{3}{c}{$P$}    & Remarks\\
\multicolumn{1}{c}{ID} & Field   &\multicolumn{1}{c}{No}& [2000.0] & [2000.0] & [\arcs] & [mag] & [mag] & \multicolumn{3}{c}{[days]} &\\   
\noalign{\vskip3pt}
\hline
\noalign{\vskip3pt}
       858 & BLG654.24 &  63566 & 17\uph33\upm26\zdot\ups90 & $-29\arcd42\arcm11\zdot\arcs7$ & 0.27 & 15.403 & --   &  11&.&174    & SP\\
       869 & BLG648.13 &  45934 & 17\uph45\upm36\zdot\ups79 & $-26\arcd51\arcm32\zdot\arcs7$ & 0.31 & 14.130 & --   &  23&.&81     & SP\\
       877 & BLG504.24 & 106420 & 17\uph55\upm14\zdot\ups65 & $-27\arcd59\arcm01\zdot\arcs4$ & 0.36 & 12.763 & 2.472 &  27&.&933    & SP\\
       878 & BLG501.24 &   8191 & 17\uph50\upm01\zdot\ups72 & $-29\arcd42\arcm56\zdot\arcs7$ & 0.71 & 15.848 & 1.646 &   0&.&64616  & E\\
       895 & BLG654.23 &  73640 & 17\uph34\upm05\zdot\ups41 & $-29\arcd42\arcm43\zdot\arcs8$ & 0.81 & 15.979 & --   &   0&.&42973  & E\\
       896 & BLG645.03 &  63999 & 17\uph58\upm12\zdot\ups28 & $-27\arcd21\arcm39\zdot\arcs3$ & 0.77 & 13.707 & --   &   3&.&8534   & E\\
       898 & BLG504.29 &  96872 & 17\uph57\upm28\zdot\ups81 & $-27\arcd33\arcm10\zdot\arcs7$ & 0.60 & 14.798 & 3.099 &  44&.&84     & SP\\
       899 & BLG504.29 & 118157 & 17\uph57\upm18\zdot\ups94 & $-27\arcd36\arcm08\zdot\arcs7$ & 0.34 & 12.976 & 3.020 &  49&.&751    & SP\\
       901 & BLG504.30 &  59047 & 17\uph56\upm51\zdot\ups19 & $-27\arcd23\arcm50\zdot\arcs6$ & 0.75 & 13.094 & 0.829 &   0&.&39967  & E\\
       906 & BLG504.31 &  26678 & 17\uph56\upm16\zdot\ups81 & $-27\arcd26\arcm34\zdot\arcs2$ & 0.55 & 13.798 & 2.126 &   0&.&29347  & SP\\
       907 & BLG504.23 &  59042 & 17\uph56\upm11\zdot\ups47 & $-27\arcd52\arcm33\zdot\arcs5$ & 3.43 & 15.710 & --   & 172&.&4      & SRV\\
       908 & BLG504.31 &  22399 & 17\uph56\upm10\zdot\ups29 & $-27\arcd32\arcm58\zdot\arcs1$ & 1.08 & 17.148 & --   &  67&.&1      & SRV\\
       909 & BLG504.31 &  20360 & 17\uph56\upm09\zdot\ups94 & $-27\arcd35\arcm17\zdot\arcs5$ & 0.42 & 15.969 & 2.516 &   4&.&0967   & SP\\
       910 & BLG504.23 &  90667 & 17\uph56\upm07\zdot\ups52 & $-27\arcd57\arcm26\zdot\arcs4$ & 0.23 & 15.288 & 1.947 &   1&.&24425  & SP\\
       912 & BLG504.31 &  42181 & 17\uph56\upm03\zdot\ups86 & $-27\arcd28\arcm33\zdot\arcs6$ & 0.62 & 15.458 & 2.346 &   0&.&37699  & P\\
       914 & BLG504.31 &  33921 & 17\uph55\upm59\zdot\ups55 & $-27\arcd37\arcm44\zdot\arcs9$ & 0.65 & 13.769 & 0.833 &  --&&        & I\\
       915 & BLG504.31 &  58984 & 17\uph55\upm58\zdot\ups61 & $-27\arcd29\arcm56\zdot\arcs8$ & 2.29 & 16.630 & 1.647 &   1&.&1609   & P\\
       924 & BLG500.17 &  36873 & 17\uph55\upm00\zdot\ups48 & $-28\arcd19\arcm56\zdot\arcs5$ & 0.20 & 13.777 & 1.991 &  77&.&5      & P\\
       926 & BLG504.16 &  57832 & 17\uph54\upm49\zdot\ups47 & $-28\arcd07\arcm36\zdot\arcs6$ & 0.42 & 17.195 & 2.210 &   3&.&787    & P\\
       929 & BLG500.17 & 144211 & 17\uph54\upm30\zdot\ups83 & $-28\arcd33\arcm35\zdot\arcs5$ & 0.19 & 16.227 & 1.695 &   0&.&59379  & SP\\
       930 & BLG500.17 & 151398 & 17\uph54\upm29\zdot\ups31 & $-28\arcd31\arcm52\zdot\arcs8$ & 2.99 & 16.716 & 1.665 &  --&&        & I\\
       933 & BLG500.09 & 120810 & 17\uph54\upm05\zdot\ups77 & $-28\arcd48\arcm23\zdot\arcs4$ & 2.27 & 15.607 & 3.306 &  --&&        & I\\
       938 & BLG500.09 & 190630 & 17\uph53\upm52\zdot\ups05 & $-28\arcd44\arcm52\zdot\arcs7$ & 1.33 & 13.735 & 0.786 &   2&.&42096  & E\\
       940 & BLG500.19 &  52204 & 17\uph53\upm30\zdot\ups09 & $-28\arcd32\arcm47\zdot\arcs7$ & 0.39 & 15.330 & 1.337 &  --&&        & I\\
       944 & BLG500.28 &     13 & 17\uph52\upm54\zdot\ups34 & $-28\arcd16\arcm29\zdot\arcs9$ & 2.48 & 14.334 & 5.269 &  42&.&02     & SRV\\
       946 & BLG500.11 &  55525 & 17\uph52\upm52\zdot\ups63 & $-28\arcd48\arcm05\zdot\arcs4$ & 3.22 & 14.142 & 1.975 &  15&.&03759  & SP\\
       947 & BLG500.03 &  89767 & 17\uph52\upm45\zdot\ups16 & $-28\arcd59\arcm15\zdot\arcs1$ & 3.42 & 15.918 & 5.493 &  14&.&881    & OSARG\\
       948 & BLG500.03 & 100501 & 17\uph52\upm41\zdot\ups86 & $-29\arcd12\arcm52\zdot\arcs6$ & 0.36 & 14.925 & 1.430 &   1&.&9577   & SP\\
       953 & BLG500.21 &  34612 & 17\uph52\upm03\zdot\ups78 & $-28\arcd33\arcm51\zdot\arcs1$ & 0.34 & 15.901 & 1.534 &   1&.&0989   & SP\\
       955 & BLG500.12 & 106387 & 17\uph51\upm51\zdot\ups42 & $-28\arcd42\arcm46\zdot\arcs5$ & 0.35 & 16.386 & --   &  --&&        & I\\
{\it   964}& {\it BUL\_SC5} & {\it 172053} & {\it 17\uph50\upm15\zdot\ups31} & ${\it -29\arcd39\arcm13\zdot\arcs8}$ & {\it 0.19} & {\it 12.289} & {\it 1.681} &  {\it 81}&.&{\it 97} & {\it P}\\
       966 & BLG501.15 &  49578 & 17\uph49\upm42\zdot\ups67 & $-30\arcd02\arcm44\zdot\arcs5$ & 1.96 & 16.881 & 3.170 &   9&.&662    & SP\\
       969 & BLG501.15 &  71683 & 17\uph49\upm32\zdot\ups21 & $-30\arcd03\arcm50\zdot\arcs4$ & 0.60 & 18.335 & --   &  32&.&154    & P\\
       970 & BLG501.24 &  80876 & 17\uph49\upm28\zdot\ups07 & $-29\arcd46\arcm29\zdot\arcs1$ & 0.70 & 12.504 & 0.970 &  --&&        & I\\
       972 & BLG534.32 & 100700 & 17\uph49\upm23\zdot\ups67 & $-30\arcd32\arcm16\zdot\arcs2$ & 0.30 & 12.755 & 0.921 &   0&.&334094 & E\\
       973 & BLG501.16 &     69 & 17\uph49\upm21\zdot\ups61 & $-30\arcd07\arcm39\zdot\arcs2$ & 0.45 & 16.672 & 2.838 &  10&.&352    & SP\\
       978 & BLG501.25 &  42282 & 17\uph48\upm59\zdot\ups54 & $-29\arcd39\arcm40\zdot\arcs0$ & 2.32 & 16.675 & 1.708 &   1&.&05765  & P\\
       981 & BLG501.16 &  49689 & 17\uph48\upm54\zdot\ups94 & $-29\arcd54\arcm16\zdot\arcs7$ & 0.51 & 15.823 & 1.335 &  91&.&7      & P\\
      1019 & BLG648.13 &  46396 & 17\uph45\upm48\zdot\ups53 & $-26\arcd50\arcm50\zdot\arcs5$ & 0.18 & 16.406 & --   &   0&.&35143  & E\\
      1033 & BLG648.31 &  60638 & 17\uph45\upm02\zdot\ups19 & $-26\arcd14\arcm49\zdot\arcs4$ & 0.23 & 13.104 & --   &  54&.&07799  & SP\\
      1070 & BLG675.26 &  56707 & 17\uph42\upm58\zdot\ups73 & $-26\arcd55\arcm01\zdot\arcs4$ & 0.39 & 14.560 & --   &  47&.&17     & SP\\
      1071 & BLG675.01 &  49220 & 17\uph42\upm57\zdot\ups12 & $-27\arcd46\arcm30\zdot\arcs2$ & 0.55 & 15.767 & --   &  37&.&175    & P\\
      1074 & BLG675.26 &  74928 & 17\uph42\upm52\zdot\ups61 & $-26\arcd57\arcm14\zdot\arcs1$ & 0.96 & 15.156 & --   &   3&.&1616   & E\\
      1078 & BLG675.19 &  37341 & 17\uph42\upm24\zdot\ups03 & $-27\arcd11\arcm57\zdot\arcs8$ & 2.61 & 16.536 & --   &  --&&        & I\\
      1086 & BLG675.03 &  34167 & 17\uph41\upm46\zdot\ups47 & $-27\arcd49\arcm54\zdot\arcs7$ & 0.03 & 16.526 & --   &  11&.&768    & E\\
      1090 & BLG675.11 & 107190 & 17\uph41\upm25\zdot\ups07 & $-27\arcd31\arcm43\zdot\arcs6$ & 0.67 & 14.827 & --   &  49&.&26     & SP\\
      1104 & BLG675.21 & 104241 & 17\uph40\upm40\zdot\ups67 & $-27\arcd08\arcm07\zdot\arcs9$ & 3.88 & 15.119 & --   & 109&.&89     & P\\
      1110 & BLG675.13 &  40278 & 17\uph40\upm17\zdot\ups69 & $-27\arcd30\arcm57\zdot\arcs1$ & 0.44 & 13.617 & --   &  30&.&612    & E\\
      1112 & BLG675.22 &  79334 & 17\uph40\upm12\zdot\ups92 & $-27\arcd12\arcm35\zdot\arcs7$ & 1.26 & 14.275 & --   &  17&.&699    & SP\\
      1151 & BLG653.01 &   1611 & 17\uph37\upm55\zdot\ups27 & $-29\arcd10\arcm10\zdot\arcs1$ & 2.67 & 14.616 & --   &  78&.&125    & SRV\\
{\bf 1155} & BLG654.26 &  59596 & 17\uph37\upm31\zdot\ups10 & $-29\arcd24\arcm28\zdot\arcs8$ & 0.38 & 13.476 & --   &   0&.&6462   & E\\
{\bf 1155} & BLG654.26 &  59618 & 17\uph37\upm31\zdot\ups33 & $-29\arcd24\arcm28\zdot\arcs9$ & 3.35 & 15.531 & --   &  56&.&8      & SRV\\
\noalign{\vskip3pt}
\hline}
\setcounter{table}{0}
\MakeTable{r@{\hspace{7pt}}c@{\hspace{3pt}}r@{\hspace{7pt}}c@{\hspace{7pt}}c@{\hspace{7pt}}c@{\hspace{7pt}}c@{\hspace{7pt}}c@{\hspace{5pt}}r@{\hspace{0pt}}c@{\hspace{0pt}}l@{\hspace{5pt}}l}
{12.5cm}{Concluded}
{\hline
\noalign{\vskip3pt}
     GBS               & OGLE-IV & OGLE-IV              &    RA    &   DEC    & $D$     & $I$   & $V-I$ & \multicolumn{3}{c}{$P$}    & Remarks\\
\multicolumn{1}{c}{ID} & Field   &\multicolumn{1}{c}{No}& [2000.0] & [2000.0] & [\arcs] & [mag] & [mag] & \multicolumn{3}{c}{[days]} &\\   
\noalign{\vskip3pt}
\hline
      1161 & BLG653.19 &  67406 & 17\uph36\upm45\zdot\ups56 & $-28\arcd34\arcm14\zdot\arcs8$ & 3.15 & 15.511 & --   &  94&.&3      & SRV\\
      1165 & BLG653.20 &  10600 & 17\uph36\upm33\zdot\ups67 & $-28\arcd30\arcm03\zdot\arcs9$ & 0.14 & 13.703 & --   &  74&.&1      & SRV\\
      1168 & BLG654.03 &  26666 & 17\uph36\upm27\zdot\ups84 & $-30\arcd12\arcm14\zdot\arcs1$ & 0.52 & 13.823 & --   &  65&.&79     & P\\
      1179 & BLG654.11 &  52361 & 17\uph36\upm06\zdot\ups21 & $-29\arcd55\arcm40\zdot\arcs5$ & 2.14 & 15.369 & --   &  --&&        & I\\
      1185 & BLG653.12 &   9604 & 17\uph35\upm48\zdot\ups65 & $-28\arcd47\arcm09\zdot\arcs7$ & 0.94 & 14.880 & --   &   2&.&1497   & E\\
      1187 & BLG653.04 &   7005 & 17\uph35\upm45\zdot\ups73 & $-29\arcd08\arcm02\zdot\arcs7$ & 2.71 & 15.012 & --   &  16&.&92     & OSARG\\
      1194 & BLG654.04 &  62415 & 17\uph35\upm27\zdot\ups50 & $-30\arcd14\arcm01\zdot\arcs3$ & 2.51 & 16.015 & --   &   1&.&91022  & SP\\
      1208 & BLG654.06 &  16127 & 17\uph34\upm23\zdot\ups71 & $-30\arcd17\arcm05\zdot\arcs1$ & 0.90 & 14.935 & --   &  24&.&876    & SP\\
      1228 & BLG500.03 & 188094 & 17\uph52\upm31\zdot\ups10 & $-28\arcd58\arcm03\zdot\arcs9$ & 1.58 & 14.929 & 2.458 &   3&.&4235   & P\\
      1229 & BLG500.12 & 106361 & 17\uph51\upm56\zdot\ups83 & $-28\arcd41\arcm45\zdot\arcs7$ & 0.34 & 14.649 & 1.770 &  --&&        & I\\
\noalign{\vskip3pt}
\noalign{\vskip3pt}
\hline
\noalign{\vskip3pt}
\multicolumn{12}{p{12.5cm}}{
$\bullet$ Italic font entry highlights the counterpart candidates selected
from the OGLE-II databases.\newline
$\bullet$ Boldface GBS number highlights the cases when two variable stars are
located in the search region.\newline
$\bullet$ Variability type abbreviations: SP~--~spotted star,
OSARG/SRV/MIRA -- pulsating giant, E~--~eclipsing
system, P -- periodic system,
RRLYR~--~RR~Lyr pulsating star, CV~--~cataclysmic variable, 
I~--~irregular star, AGN -- active galactic nucleus.\newline
$\bullet$ The same optical counterpart candidates 
for two close GBS X-ray sources: GBS \#16/GBS \#515, GBS \#303/GBS \#304,
GBS \#483/GBS \#1155, GBS \#501/GBS \#946.
}
}
It should be noted that in the case of five X-ray sources from the GBS
catalog more than one variable star were found within our search radius.
The best example here is the X-ray source GBS \#19 (Fig.~7). A variable red
giant and fainter eclipsing system are detected very close to the X-ray
position. Only further, preferably spectroscopic, observations may solve
the cross-identification ambiguity in this case.  However, it may turn out
that both these stars are X-ray sources. A one day long flare of $\approx
1$~mag amplitude observed on ${\rm HJD=2\,455\,349.7}$ and ${\rm
HJD=2\,456\,071.7}$ in the eclipsing star may indicate stellar activity. On
the other hand the light curve of the red giant resembles that of other
X-ray binary systems.

There are also four cases where two separate X-rays sources from the GBS
catalog point very close to each other (\eg\ \#16 and \#515 are separated by
only 3\zdot\arcs3). In such cases the same optical variable stars are
selected as potential counterparts to these X-ray sources. Additional
observations can resolve the issue of the real connections of the sources
in such cases.

\begin{figure}[htb]
\centerline{\includegraphics[width=8.2cm, bb=110 50 510 310]{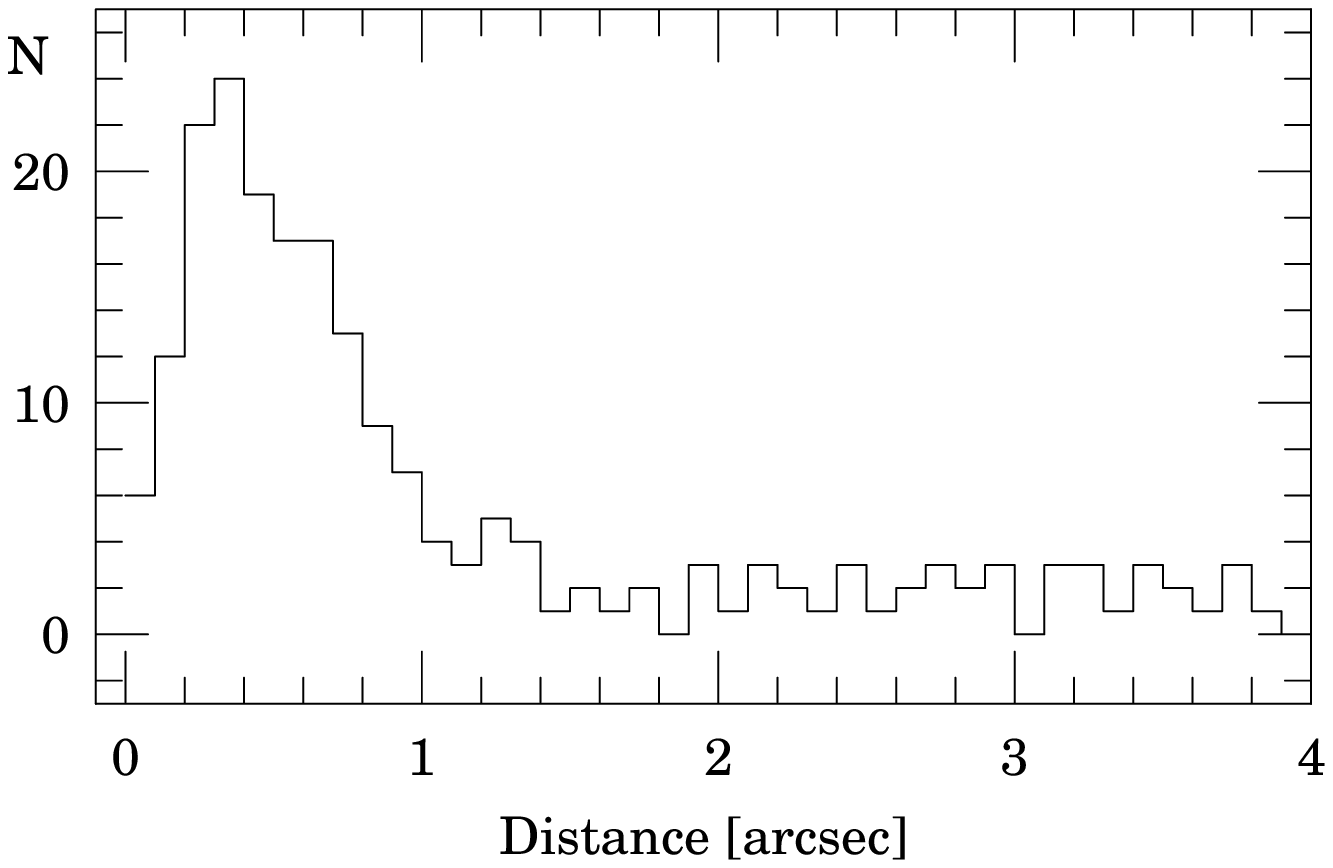}}
\FigCap{Histogram of the angular distances in the sky between the OGLE
and X-ray position of the optical counterpart candidates. The bin size
is 0\zdot\arcs1.}
\end{figure}
Fig.~8 shows the histogram of the angular distances between the OGLE
equatorial coordinates of the counterpart candidates and their X-ray
positions in 0\zdot\arcs1 bins. It is clearly seen that for the vast
majority of objects the coincidence of positions is better than 1\arcs.
In the remaining cases it is possible that the X-ray coordinates are
less accurate or a variable optical candidate is aligned by chance on
the sky with the corresponding X-ray source. Conservative selection of
the 15 pixel (3\zdot\arcs9) radius ensures high completeness of
our search.

\MakeTable{rcrccccc}{12.5cm}{Possible bright non-variable optical counterparts to 
the X-ray sources detected by the Galactic Bulge Survey}
{\hline
\noalign{\vskip3pt}
 GBS &  OGLE-IV  &OGLE-IV &  RA      & DEC      &  $D$    & $I$   & $V-I$ \\
 ID  &  Field    &\multicolumn{1}{c}{No} & [2000.0] & [2000.0] & [\arcs] & [mag] & [mag]\\
\noalign{\vskip3pt}
\hline
\noalign{\vskip3pt}
  88 & BLG654.24 &  96086 & 17\uph33\upm08\zdot\ups04 & $-29\arcd37\arcm52\zdot\arcs7$ & 0.36 & 15.423 & --\\
 149 & BLG675.27 &  31995 & 17\uph42\upm26\zdot\ups78 & $-26\arcd47\arcm31\zdot\arcs3$ & 0.57 & 12.636 & --\\
 170 & BLG654.18 &  34390 & 17\uph37\upm55\zdot\ups18 & $-29\arcd35\arcm40\zdot\arcs3$ & 0.13 & 12.980 & --\\
 294 & BLG504.25 &  67635 & 17\uph54\upm35\zdot\ups37 & $-27\arcd44\arcm59\zdot\arcs4$ & 0.49 & 14.041 & 1.783\\
 393 & BLG654.29 &  66612 & 17\uph35\upm30\zdot\ups83 & $-29\arcd18\arcm31\zdot\arcs9$ & 0.23 & 14.145 & --\\
 403 & BLG500.02 & 184262 & 17\uph53\upm06\zdot\ups53 & $-29\arcd05\arcm30\zdot\arcs1$ & 0.33 & 14.384 & 1.055\\
 454 & BLG648.23 &   8019 & 17\uph45\upm25\zdot\ups15 & $-26\arcd39\arcm08\zdot\arcs0$ & 0.49 & 12.490 & --\\
 631 & BLG653.05 &  53272 & 17\uph34\upm43\zdot\ups92 & $-29\arcd09\arcm49\zdot\arcs7$ & 0.60 & 13.521 & --\\
 646 & BLG654.30 &  12088 & 17\uph35\upm06\zdot\ups17 & $-29\arcd21\arcm12\zdot\arcs1$ & 0.36 & 13.802 & --\\
 662 & BLG504.22 & 107804 & 17\uph56\upm42\zdot\ups18 & $-27\arcd45\arcm54\zdot\arcs3$ & 0.10 & 13.144 & 1.859\\
 675 & BLG500.17 & 150996 & 17\uph54\upm34\zdot\ups00 & $-28\arcd32\arcm19\zdot\arcs7$ & 0.36 & 13.360 & 0.906\\
 773 & BLG648.16 &  12494 & 17\uph44\upm01\zdot\ups99 & $-26\arcd50\arcm49\zdot\arcs7$ & 0.75 & 13.285 & --\\
 880 & BLG501.16 &  47309 & 17\uph49\upm00\zdot\ups89 & $-29\arcd54\arcm32\zdot\arcs8$ & 0.22 & 13.102 & 0.794\\
 967 & BLG534.32 &  76738 & 17\uph49\upm37\zdot\ups15 & $-30\arcd32\arcm57\zdot\arcs8$ & 0.34 & 13.074 & 0.719\\
1045 & BLG648.32 &  45146 & 17\uph44\upm29\zdot\ups95 & $-26\arcd19\arcm49\zdot\arcs5$ & 0.29 & 15.861 & --\\
1049 & BLG648.16 &   1014 & 17\uph44\upm08\zdot\ups94 & $-26\arcd59\arcm58\zdot\arcs6$ & 0.35 & 15.554 & --\\
1123 & BLG675.06 &  51526 & 17\uph39\upm38\zdot\ups91 & $-27\arcd43\arcm56\zdot\arcs5$ & 0.15 & 16.286 & --\\
1200 & BLG654.22 &  10352 & 17\uph35\upm13\zdot\ups17 & $-29\arcd40\arcm23\zdot\arcs8$ & 0.27 & 14.655 & --\\
1209 & BLG654.31 &  13876 & 17\uph34\upm23\zdot\ups62 & $-29\arcd21\arcm55\zdot\arcs0$ & 0.67 & 13.471 & --\\
\noalign{\vskip3pt}
\hline}
Finally, it is worth noticing that in the case of 19 X-ray sources where we
did not find any optically variable objects in their neighborhood, the
position of the X-ray source points with a high accuracy -- better than
0\zdot\arcs75 -- to a well separated star brighter than
16.5~mag. Therefore, it is very likely that these stars may also be
potential optical counterparts. As can be seen from the light curves in the
OGLE X-ray optical counterparts monitoring system, XROM (Udalski 2008),
many optical counterparts to the X-ray sources can be optically quiet for
years. Table~2 lists these potential non-variable optical counterparts to
the X-ray sources.

Table~1 can be a good starting point for further follow-up observations
of the selected optical counterparts. In particular extracting new
low-mass and massive X-ray binaries would be of great importance for
further studies of these objects in the Galactic center environment. We
plan to include a subsample of the most interesting objects presented
here to the list of objects monitored regularly by the OGLE XROM system
(Udalski 2008). This can facilitate follow-up observations providing
information on the current optical behavior of the observed objects.

\Section{Data Availability}
The OGLE-IV photometric data presented in this paper are available to the
astronomical community from the OGLE Internet Archive:

\begin{center}
{\it http://ogle.astrouw.edu.pl}\\
{\it ftp://ftp.astrouw.edu.pl/ogle4/XRAY-GBS/}
\end{center}

Usage of the data is allowed under the proper acknowledgment to the OGLE
project.

\Acknow{The OGLE project has received funding from the European Research
Council under the European Community's Seventh Framework Programme
(FP7/2007-2013)/ERC grant agreement No.~246678 to AU.}

\end{document}